\documentclass[apj]{emulateapj}
\usepackage{hyperref}
\hypersetup{colorlinks=true,citecolor=blue}
\usepackage[usenames]{color}
\usepackage{natbib}
\usepackage{amsmath}
\newcommand{\Msol}{\mbox{$M_\odot$}} 
 
\newcommand{\kms}{\mbox{km s$^{-1}$}}

\usepackage[usenames]{color}
\usepackage{multirow}

\slugcomment{Draft version June 1, 2019}

\shorttitle{Star Formation Law Along the Dust Lane of Centaurus\,A }
\shortauthors{D. Espada et al.}

\begin{document}


\title{Star Formation Efficiencies at Giant Molecular Cloud Scales in the Molecular Disk of the Elliptical Galaxy NGC\,5128 (Centaurus\,A)}


\author{D. Espada}
\affil{National Astronomical Observatory of Japan, 2-21-1 Osawa, Mitaka, Tokyo 181-8588, Japan}
\affil{The Graduate University for Advanced Studies (SOKENDAI), 2-21-1 Osawa, Mitaka, Tokyo, 181-0015, Japan}
\email{daniel.espada@nao.ac.jp}

\author{S. Verley}
\affil{Departamento de F{\'i}sica Te{\'o}rica y del Cosmos, Universidad de Granada, E-18010 Granada, Spain}
\affil{Instituto Universitario Carlos I de F{\'i}sica Te{\'o}rica y Computacional, Facultad de Ciencias, E-18071 Granada, Spain}

\author{R. E. Miura}
\affil{National Astronomical Observatory of Japan, 2-21-1 Osawa, Mitaka, Tokyo 181-8588, Japan}

\author{F. P. Israel}
\affil{Sterrewacht Leiden, Leiden University, PO Box 9513, 2300 RA, Leiden, The Netherlands}

\author{C. Henkel}
\affil{Max-Planck-Institut f{\"u}r Radioastronomie, Auf dem H{\"u}gel 69, 53121, Bonn, Germany}
\affil{Dept. of Astronomy, King Abdulaziz University, PO Box 80203, 21589 Jeddah, Saudi Arabia}

\author{S. Matsushita}
\affil{Academia Sinica Institute of Astronomy and Astrophysics, 11F of Astro-Math Bldg, AS/NTU, No.1, Section 4, Roosevelt Rd, Taipei 10617, Taiwan, Republic of China}

\author{B. Vila-Vilaro}
\affil{Joint ALMA Observatory, Alonso de C{\'o}rdova, 3107, Vitacura, Santiago 763-0355, Chile}
\affil{European Southern Observatory, Alonso de C{\'o}rdova 3107, Vitacura, Santiago, Chile}

\author{J. Ott}
\affil{National Radio Astronomy Observatory, PO Box O, 1003 Lopezville Road, Socorro, NM 87801, USA}

\author{K. Morokuma-Matsui}
\affil{Institute of Space and Astronautical Science, Japan Aerospace Exploration Agency, 3-1-1 Yoshinodai, Chuo-ku, Sagamihara, Kanagawa 252-5210, Japan}
\affil{Institute of Astronomy, School of Science, The University of Tokyo, 2-21-1 Osawa, Mitaka, Tokyo 181-0015, Japan}

\author{A. B. Peck}
\affil{Gemini Observatory, 670 N'Aohoku Pl, Hilo 96720-2700, Hawaii, HI, USA}

\author{A. Hirota}
\affil{Joint ALMA Observatory, Alonso de C{\'o}rdova, 3107, Vitacura, Santiago 763-0355, Chile}
\affil{National Astronomical Observatory of Japan, 2-21-1 Osawa, Mitaka, Tokyo 181-8588, Japan}

\author{S. Aalto}
\affil{Dep. of Space, Earth and Environment, Chalmers University of Technology, Onsala Space Observatory, SE-43992 Onsala, Sweden}

\author{A. C. Quillen}
\affil{Department of Physics and Astronomy, University of Rochester, Rochester, NY 14627, USA}

\author{M. R. Hogerheijde}
\affil{Leiden Observatory, Leiden University, PO Box 9513, 2300 RA Leiden, The Netherlands}
\affil{Anton Pannekoek Institute for Astronomy, University of Amsterdam, Science Park 904, 1098 XH, Amsterdam, The Netherlands}

\author{N. Neumayer}
\affil{Max Planck Institute for Astronomy (MPIA), K{\"o}nigstuhl 17, D-69121 Heidelberg, Germany}

\author{C. Vlahakis}
\affil{National Radio Astronomy Observatory, 520 Edgemont Road, Charlottesville, VA 22903-2475, USA}

\author{D. Iono}
\affil{National Astronomical Observatory of Japan, 2-21-1 Osawa, Mitaka, Tokyo 181-8588, Japan}
\affil{The Graduate University for Advanced Studies (SOKENDAI), 2-21-1 Osawa, Mitaka, Tokyo, 181-0015, Japan}

\and

\author{K. Kohno}
\affil{Institute of Astronomy, School of Science, The University of Tokyo, 2-21-1 Osawa, Mitaka, Tokyo 181-0015, Japan}

\begin{abstract}

We present  ALMA CO(1--0) observations  toward the dust lane of the nearest elliptical and radio galaxy, NGC\,5128 (Centaurus~A), with high angular resolution ($\sim1\arcsec$, or 18\,pc), including information from large to small spatial scales and total flux. We find a total molecular gas mass of 1.6$\times$10$^9$\,$M_\odot$ and we reveal the presence of filamentary components more extended than previously seen, up to a radius of 4\,kpc. We find that the global star formation rate is $\sim$1\,\Msol\,yr$^{-1}$, which yields a star formation efficiency (SFE) of 0.6\,Gyr$^{-1}$ (depletion time $\tau =$\,1.5\,Gyr), similar to those in disk galaxies. We show the most detailed view to date (40\,pc resolution) of the relation between molecular gas and star formation within the stellar component of an elliptical galaxy, from several kpc scale to the circumnuclear region close to the powerful radio jet. Although on average the SFEs are similar to those of spiral galaxies, the circumnuclear disk (CND) presents SFEs of 0.3\,Gyr$^{-1}$, lower by a factor of 4 than the outer disk.  The low SFE in the CND is in contrast to the high SFEs found in the literature for the circumnuclear regions of some nearby disk galaxies with nuclear activity, probably as a result of larger shear motions and longer AGN feedback. The higher SFEs in the outer disk suggests that only central molecular gas or filaments with sufficient density and strong shear motions will remain in $\sim$1\,Gyr, which will later result in the compact molecular distributions and low SFEs usually seen in other giant ellipticals with cold gas. 

\end{abstract}


\section{Introduction}

Cold molecular clouds are the sites of star formation (SF) in galaxies and their relation under various galactic environments is essential to understand galaxy evolution. The relation is usually expressed in the form of the so called Kennicutt-Schmidt (KS) SF law, i.e. the correlation between star formation rate (SFR) and molecular gas content \citep{1959ApJ...129..243S,1998ARA&A..36..189K,2008AJ....136.2846B}. Although this relation holds for several orders of magnitude,
other processes may break the KS law at scales smaller than giant molecular cloud scales, such as star forming activities \citep[e.g.][]{2010ApJ...722L.127O,2014ApJ...788..167M}. Also, at larger scales, different SF laws are present depending on the specific properties of the environment. Starburst galaxies are known to be characterized by much higher star formation efficiencies (SFEs), or equivalently lower depletion times, than other local disk galaxies \citep[e.g.][]{2010ApJ...714L.118D}. The central regions of a sample of four spiral galaxies possessing low luminosity Active Galactic Nuclei (AGN) are characterized by higher SFEs than in other regions \citep{2015A&A...577A.135C}.

However, the KS SF laws in elliptical galaxies are too poorly known to infer the fate of their molecular gas and to compare it with the properties of the more widely studied disk galaxies. 
Molecular gas has been detected in a significant fraction of early type galaxies \citep{2014MNRAS.444.3408Y}, and 
recent (less than 1\,Gyr ago) SF has also been detected in $\sim$ 20\% of the early type galaxies \citep[e.g.][]{2005ApJ...619L.111Y}.
Nevertheless, high angular resolution studies in elliptical galaxies are hampered by a lack of objects that are located nearby and 
samples of early type galaxies usually mix ellipticals with a large number of lenticular galaxies. 
Furthermore,  significantly less molecular gas exists in elliptical galaxies, and it is more centrally concentrated than in spiral galaxies of comparable total mass \citep{2010ApJ...725L..62W}.
Regarding SF activities, conflicting results exist in the literature.
The derived SFR and molecular gas surface densities place the E/S0 galaxies overlapping the range spanned by the disks and centers of spiral galaxies and the relatively constant efficiency SF laws, although with a larger scatter \citep[e.g.][]{2010ApJ...725L..62W,2011MNRAS.410.1197C,2017A&A...605A..74K}, while others suggest that the SF in early type galaxies is suppressed and the corresponding SF surface densities lie below the standard KS relation of late-type galaxies \citep{2014MNRAS.444.3427D,2015MNRAS.449.3503D,2018MNRAS.476..122V}.

Besides these conflicting results, it has not been possible to resolve spatially, and with sufficient signal to noise ratio (S/N), the molecular and SF properties of a large number of molecular cloud samples within giant elliptical galaxies to properly address the issue. The molecular gas properties have been difficult to obtain because instrumentation in the past was not able to provide high angular resolution combined with sufficient sensitivity and dynamic range. The Atacama Large Millimeter/submillimeter Array (ALMA) is now starting to provide insight into the molecular properties of truly elliptical galaxies.
Also, it is not straightforward to obtain the SFR in these objects, as the recipes used for spiral galaxies may not be valid for ellipticals.

The main goal of this paper is to provide the resolved KS SF law at giant molecular cloud (GMC) scales within the nearest giant elliptical and radio galaxy, NGC\,5128 (Cen~A), with high sensitivity from kpc scales to regions close to the powerful AGN.
Cen~A is at a distance of $D$ $\simeq$ 3.8\,Mpc (\citealt{2010PASA...27..457H}, where 1\arcsec = 18\,pc) and is thus the closest target among the class of elliptical galaxies for resolved studies of the molecular gas component and its relation to SF. Furthermore, it is also ideal to investigate the influence of its
        massive densely packed stellar body and powerful 
        nuclear activity on the molecular gas.
The gaseous component in Cen~A  is believed to have been replenished recently (a few 10$^8$\,yr) by gas from an external source, probably the accretion of an \ion{H}{1}-rich galaxy { \citep[e.g.][]{2010A&A...515A..67S}}. The recent gas accretion event allows us to study with unprecedented detail a relatively stable and settled molecular disk extending several kpc within the elliptical galaxy. This promises to shed light onto the SF activities and survival of molecular gas within an elliptical galaxy before it is consumed or destroyed. 

The dust lane along the minor axis of { Cen~A} may contain a molecular gas reservoir of 10$^{8}$ -- 10$^{9}$\,$M_\odot$ (converted to our convention of distance) as traced by several molecular transitions \citep[e.g.][]{1987ApJ...322L..73P,1990ApJ...363..451E,1993A&A...270L..13R,2001A&A...371..865L,2017ApJ...851...76M}. 
The molecular disk is also associated with other components of the interstellar medium, such as 
 ionized gas traced by H$\alpha$ (e.g. \citealt{1992ApJ...387..503N}), mainly stellar emission traced by the 
        near infrared \citep{1993ApJ...412..550Q}, and dust in the submillimeter \citep[e.g.][]{1993MNRAS.260..844H,2002ApJ...565..131L} and mid-IR continuum emission \citep[e.g.][]{1999A&A...341..667M,2006ApJ...645.1092Q}. The molecular gas distribution forms a kpc scale spiral feature \citep{2012ApJ...756L..10E}. Inside there is a $\sim$ 400\,pc sized ($\sim$ 24\arcsec)  circumnuclear disk (CND) at a position angle ($P.A.$) of $155\arcdeg$, which is perpendicular to the inner radio and X-ray jet, at least in projection \citep{2009ApJ...695..116E}.  The estimated total gas mass in this component is 9 $\times$ 10$^7$\,$M_\odot$ \citep{2014A&A...562A..96I,2017A&A...599A..53I}. The CND has been studied in detail, even with a linear resolution of $\sim$ 5\,pc (0\farcs3) in several molecular transitions. Multiple internal filaments and shocks are seen to be present likely caused by non-circular motions \citep{2017ApJ...843..136E}.

The molecular gas in both the extended disk and the CND finds its origin in the same recently accreted galaxy and the basic physical properties are expected to be similar throughout. 
A relatively constant metallicity as a function of radius of $Z$ $\simeq$ 0.75\,$Z_\odot$  was found \citep{2017A&A...599A..53I}, even in the CND, probably because the accreted gas is already well-mixed. \citet{2014ApJ...787...16P}  infer by comparing atomic cooling line and ionized gas data with photodissociation region models that the strength of the impinging far-ultraviolet radiation field in the dust lane varies from $G_0$ = 55 and 550, and the total hydrogen densities range between $n$ = 500 and 5000\,cm$^{-3}$. The molecular gas throughout Cen A is not unlike that in the disks of spiral galaxies, except for the lack of radial gradients \citep{2014ApJ...787...16P}. However, the central gas should differ from the extended disk gas insofar as it is influenced by its proximity to a major AGN. A possible consequence of the activities of the AGN is that the large scale average gas-to-dust mass ratio was found to be $\sim$100, but towards the CND $\sim$275 \citep{2012MNRAS.422.2291P}. \citet{2017A&A...599A..53I} found that the radiation field in the CND is weaker by an order of magnitude than in the extended disk, which was interpreted as a lack of SF in the former. 

The properties of molecular gas and the corresponding SF activities along the dust lane of Cen~A with sufficiently high angular resolution to resolve giant molecular clouds are largely unknown.
Although global estimates of SFRs comprising all the molecular disk exist in the literature, a more detailed analysis of the resolved SF law and SFEs is still missing.
In this paper, we report high angular resolution, dynamic range, and sensitive CO(1--0) observations obtained with ALMA covering most of the dust lane, which enable us to obtain molecular gas surface densities which can be compared with the SFR surface densities at resolutions of $\sim$ 40\,pc. 
In a companion paper we exploit the CO(1--0) data cube to build a complete census of GMCs across the molecular disk of Cen~A (\citealt{PaperII}, hereafter Paper II).

The outline of this paper is as follows. The ALMA observations, archival data compilation and
data reduction are summarized in \S\,\ref{obs}. In \S\,\ref{result} we
present the CO maps, the different molecular gas components, and derive their main properties, and in \S\,\ref{sfr} we obtain global SFR properties using Spitzer mid-IR and GALEX FUV data. 
In \S\,\ref{subsec:resolvedsflaw} we present local ($\sim$ 40\,pc resolution) molecular gas and SFR surface densities, as well as the resulting KS SF law. In \S\,\ref{discussion} we discuss our results in the context of other literature samples, highlighting the comparison with the SF law properties found in other early type galaxies.

\section{Observations and Archival Data}\label{obs}

\subsection{ALMA CO(1--0) Observations}
\label{co1-0}

\begin{figure*}
\begin{center}
\includegraphics[width=.8\textwidth]{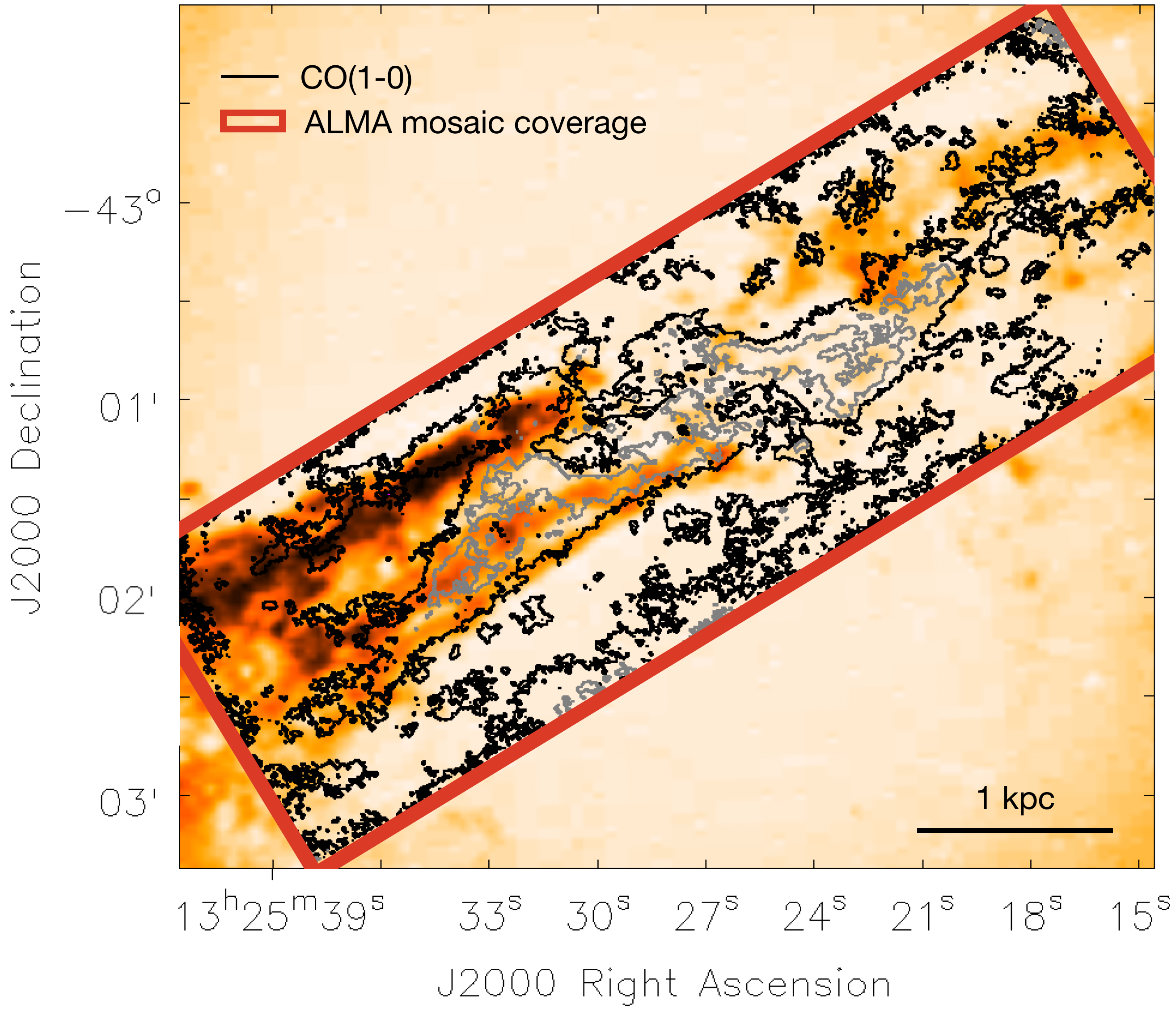}
\caption{ALMA band 3 field of view and CO(1--0) integrated intensity contours over the optical image (POSS2-Blue) of NGC\,5128 (Centaurus~A). Most of the dust lane is covered by the ALMA observations. { The contour levels are 3 (black) and 15\,$\sigma$ (gray)}, where $\sigma=0.14$\,Jy\,beam$^{-1}$\kms. The HPBW of the CO(1--0) map is $1\farcs$39 $\times$ $1\farcs$05 (P.A. = 62\arcdeg).
\label{almafov}}
\end{center}
\end{figure*}

We observed the CO(1--0) line towards a pointed mosaic covering most of the dust lane of Cen~A (see Fig.\,\ref{almafov}). 
The observations include 12m, 7m, and Total Power (TP) arrays, and thus the maps capture information from large to small spatial scales.
The mosaic's size was defined as 5\arcmin $\times$1\farcm4 and a position angle of P.A. = 120\arcdeg.
This was done with 46 pointings in the 12m array and 19 pointings in the 7m array. Nyquist sampling was used for the separation of the different pointings. The half power beam width (HPBW) is $50\farcs6$ and $86\farcs8$ at 115\,GHz 
for the 12m and 7m antennas, respectively. The TP array raster map was a rectangular field of $405\arcsec \times 189\arcsec$. 

The ALMA CO datasets were obtained during 2014 and 2015 as part of program 2013.1.00803.S (P.I. D. Espada).
The CO(1--0) dataset consisted of 10 executions: two for the 12m arrays (one extended and one compact configuration dataset), two 7m array, and six TP array datasets. 
The extended and compact 12m array configurations had longest baseline lengths of 650.3\,m  and 348.5\,m, respectively. 
The execution IDs, observation dates, time on source and calibrators used are given in Table\,\ref{tbl1}.
The total time on source for extended 12m / compact 12m / 7m and TP arrays were 27.8/13.9/57.2/116.2\,min.

The spectral setup was designed to observe the CO(1--0) line ($\nu_{\rm rest}$=115.271\,GHz). 
The spectral window in the upper sideband where the line was centered had a bandwidth of $\sim$2450\,\kms\ (937.5\,MHz) and a velocity resolution of 1.2\,\kms\  (488.281\,kHz). 
Additional spectral windows were placed at lower frequencies (centre sky frequencies 112.914, 100.999 and 102.699\,GHz) 
using bandwidths of 1875\,MHz and resolutions of 31.25\,MHz.

Data calibration and imaging were performed using the Common Astronomy Software Applications \citep[CASA;][]{2007ASPC..376..127M}. 
We used the packages containing the calibration files provided by the ALMA project.
The datasets were delivered using different CASA versions so we opted to perform the data reduction under version 5.1.1 using the scriptForPI.py script provided by ALMA.

The calibration scheme for all the datasets was standard. We carried out phase calibration (using water vapor radiometer data), system temperature calibration, as well as bandpass, phase, amplitude and absolute flux calibration using the calibrators indicated in Table\,\ref{tbl1}.
We calibrated each of the interferometric datasets independently and concatenated them after performing line-free continuum subtraction.
We checked that visibilities from different elements (extended and compact 12m, 7m, TP arrays) had correct weights after calibration.
These weights were calculated inside CASA by taking into account the system temperature ($T_{\rm sys}$) variance, integration time, and antenna diameters.

The single dish data were calibrated in units of antenna temperature ($T_a^{*}$) in Kelvin with frequent observations of blank sky (i.e. OFF position). 
Further residual baseline corrections were done using line-free channels.
A scaling factor of 41\,Jy\,K$^{-1}$ between $T_a^{*}$ and flux density (Kelvin to Jy/beam factor) as obtained by the observatory using a bright quasar for band 3 was applied to the data to convert them to units of Jy/beam.

The CO(1--0) interferometric 12m and 7m data were combined and imaged using the {\verb TCLEAN } task with Briggs weighting and robust parameter equal to 0.5. 
In this step CASA 5.4 was used as inaccuracies in the primary beam pattern have been recently found in earlier CASA versions.
The spatial resolution of the final images is $1\farcs$39 $\times$ $1\farcs$05 (24.8 $\times$ 18.9\,pc), and P.A. = 62\arcdeg. 
The resulting data cubes were then combined in CASA with the single dish data cubes using the feathering technique.
We generated CO data cubes ranging from $V_{\rm LSRK}=242$\,\kms\ to 820\,\kms\ (Kinematic Local Standard of Rest velocity) and with 2.0\,\kms\ resolution.
The combined CO(1--0) data cube is characterized by a typical noise level of 10\,mJy\,beam$^{-1}$ for a channel width of 2\,\kms .

\begin{figure}[htbp]
\begin{center}
\includegraphics[width=9cm]{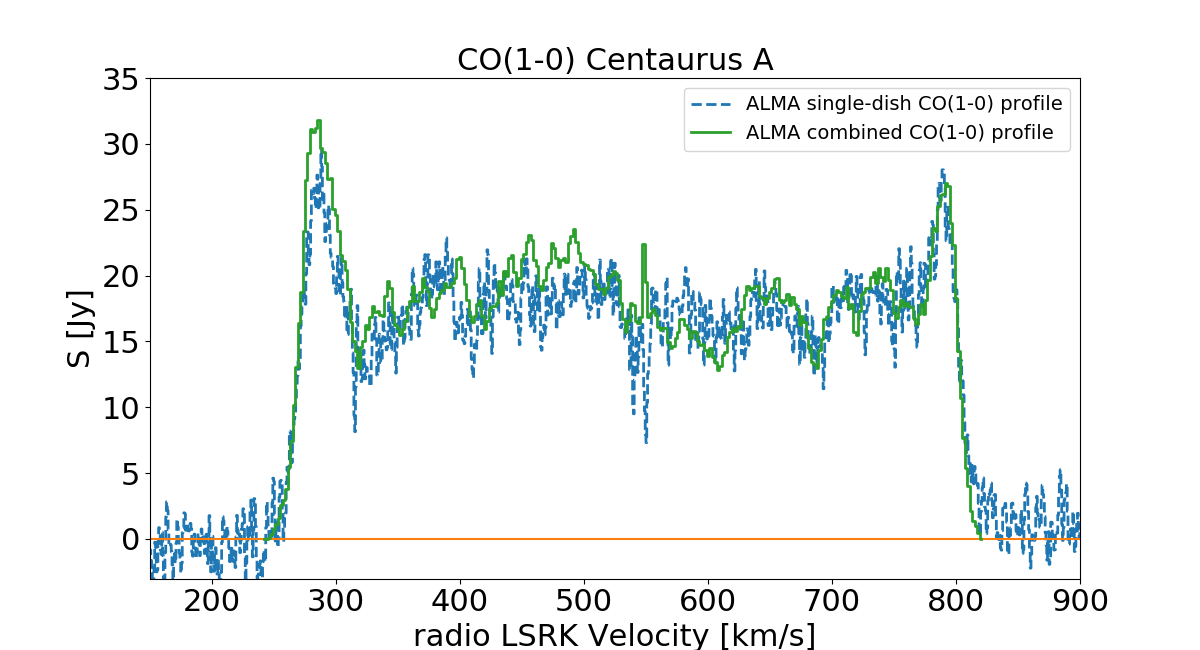}
\caption{ALMA single-dish CO(1--0) profile (green full line) compared to the profile obtained from the masked and combined (12m+7m+TP arrays) CO(1--0) data cube (blue dashed line). }
\label{co1-0spectra}
\end{center}
\end{figure}

Finally, we produced moment maps.
We first smoothed the CO(1--0)  datacubes to 20 $\times$ 20\,pixels (the size of a pixel is 0\farcs2 $\times$ 0\farcs2) in the spatial coordinates (Gaussian kernel) and 3 pixels (boxcar) in velocity. 
Then we obtained masks for regions above a 1$\sigma$ threshold in the CO(1--0)  smoothed datacube, which were then applied to the original datacube after visual inspection.  Note that at the edges of the mosaic there are regions with lower sensitivity but we exclude them from our analysis. Also, we flagged the absorption line information \citep[e.g.][]{2010ApJ...720..666E} towards the continuum emission in our analysis. Fig.\,\ref{co1-0spectra} shows the good agreement between the ALMA single-dish CO(1--0) profile and that obtained from the masked combined CO(1--0) data cube.

\subsection{Spitzer 3.6\,$\mu$m, 8\,$\mu$m, and 24\,$\mu$m Archival Data}\label{spitzerdata}

We compiled {\sl Spitzer} 3.6\,$\mu$m (emission mostly from the old stellar population), 8\,$\mu$m (polycyclic aromatic hydrocarbons -PAH, plus warm dust emission), and 24\,$\mu$m (warm dust) data from the Spitzer Heritage Archive. The 3.6\,$\mu$m and 8\,$\mu$m data were obtained with the IRAC instrument \citep[InfraRed Array Camera,][]{2004ApJS..154...10F}, and the 24\,$\mu$m data with the MIPS instrument \citep[Multiband Imaging Photometer for Spitzer,][]{2004SPIE.5487...50R} onboard the Spitzer Space Telescope \citep{2004ApJS..154....1W}. 

The IRAC 3.6\,$\mu$m and 8\,$\mu$m maps used in this paper were presented in \citet{2006ApJ...641L..29Q} and \citet{2006ApJ...645.1092Q}. The angular resolutions (Full Width at Half Maximum, FWHM) are 2\farcs2 and 2\farcs4, respectively. 
The MIPS data are post-BCD (higher-level products) processed using the MIPS Data Analysis Tool \citep{2005PASP..117..503G}. 
The FWHM of the MIPS point-spread function (PSF) at 24\,$\mu$m is 6\farcs0. To remove the backgrounds in all the maps, we first excluded the sources in each image by masking them and then used a sigma clipping technique to robustly estimate the background level which had to be subtracted.

We also removed the old-stellar emission contribution as traced by the 3.6\,$\mu$m data from the 8.0\,$\mu$m image, assuming that fluxes at 3.6\,$\mu$m trace stellar emission only. To do this we followed the recipe $L_\nu\ ({\rm PAH}) = L_\nu\ (8\,\mu{\rm m}) - 0.232 \times L_\nu\ (3.6\,\mu{\rm m})$ \citep{2004ApJS..154..253H, 2010ApJ...713..626B}.

\subsection{GALEX FUV Data}

The far ultraviolet (FUV, $1350-1750\,\AA$) emission from the young stars of Cen\,A is investigated through the {\sl Galaxy Evolution Explorer (GALEX)} \citep{2005ApJ...619L...1M} image, delivered by \citet{2007ApJS..173..185G}, where a description of the reduction procedure is given. We corrected for the Galactic extinction using the conversion factor $A_{\rm FUV} = 7.9 E(B-V)$ \citep{2007ApJS..173..185G} and the \citet{2011ApJ...737..103S} recalibration of the \citet{1998ApJ...500..525S} infrared-based dust map, assuming an $R_V = 3.1$ \citet{1999PASP..111...63F} reddening law. The image was provided with the background already subtracted and no further treatment has been applied.

\section{Molecular Gas}
\label{result}

\subsection{Molecular Components}
\begin{figure*}
\begin{center}
\includegraphics[width=16cm]{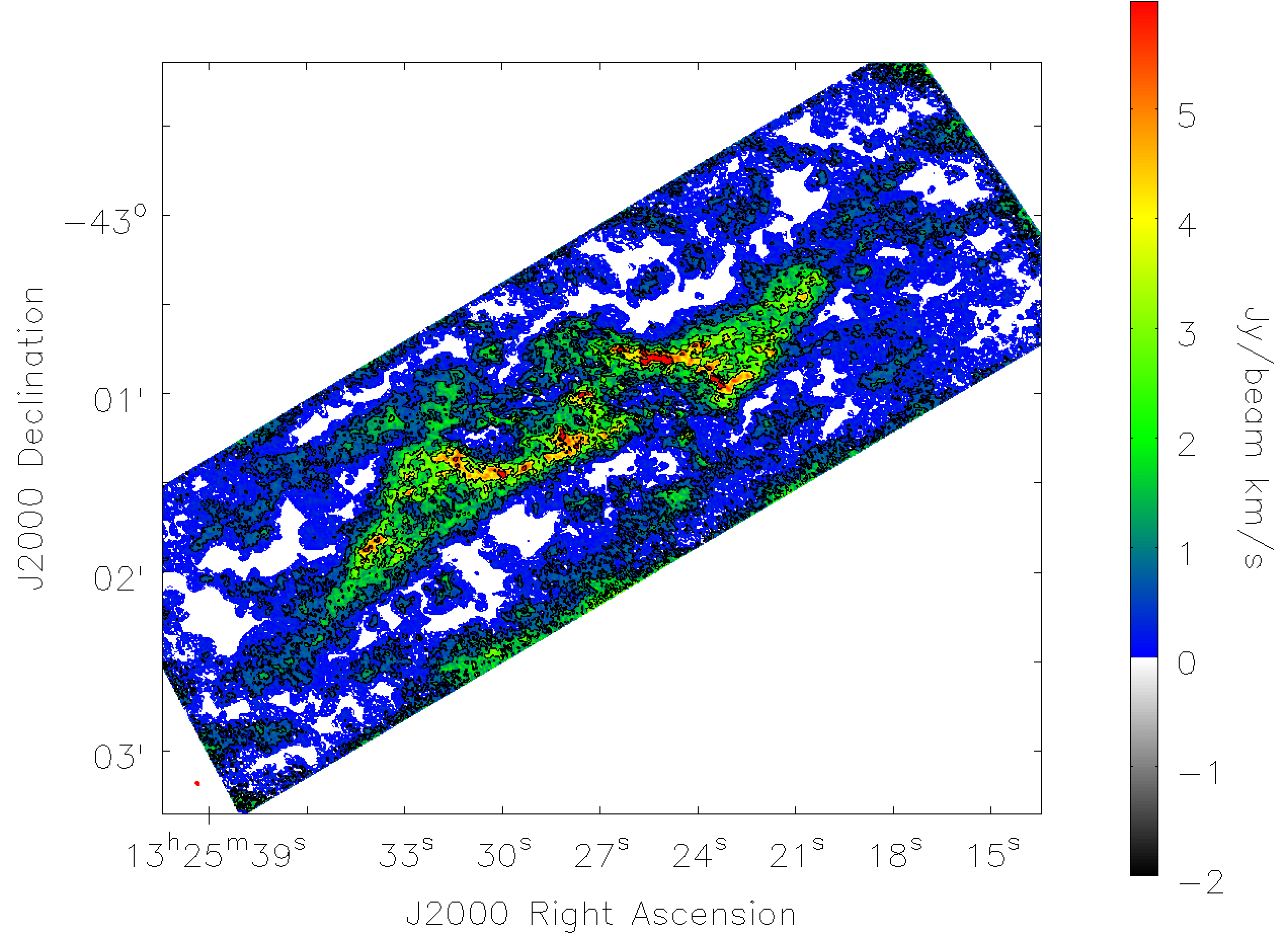}
\end{center}
\caption{
CO(1--0) integrated-intensity map of the molecular disk of Centaurus~A in color scale and contours. The contour levels are 3, 7, 15, 25, and 40\,$\sigma$, where $\sigma=0.14$\,Jy\,beam$^{-1}$\kms. The synthesized beam is shown as a red ellipse in the lower left corner. The HPBW of the CO(1--0) map is $1\farcs$39 $\times$ $1\farcs$05 (P.A. = 62\arcdeg). 
\label{mom}}
\end{figure*}

\begin{figure*}
\begin{center}
\includegraphics[width=14cm]{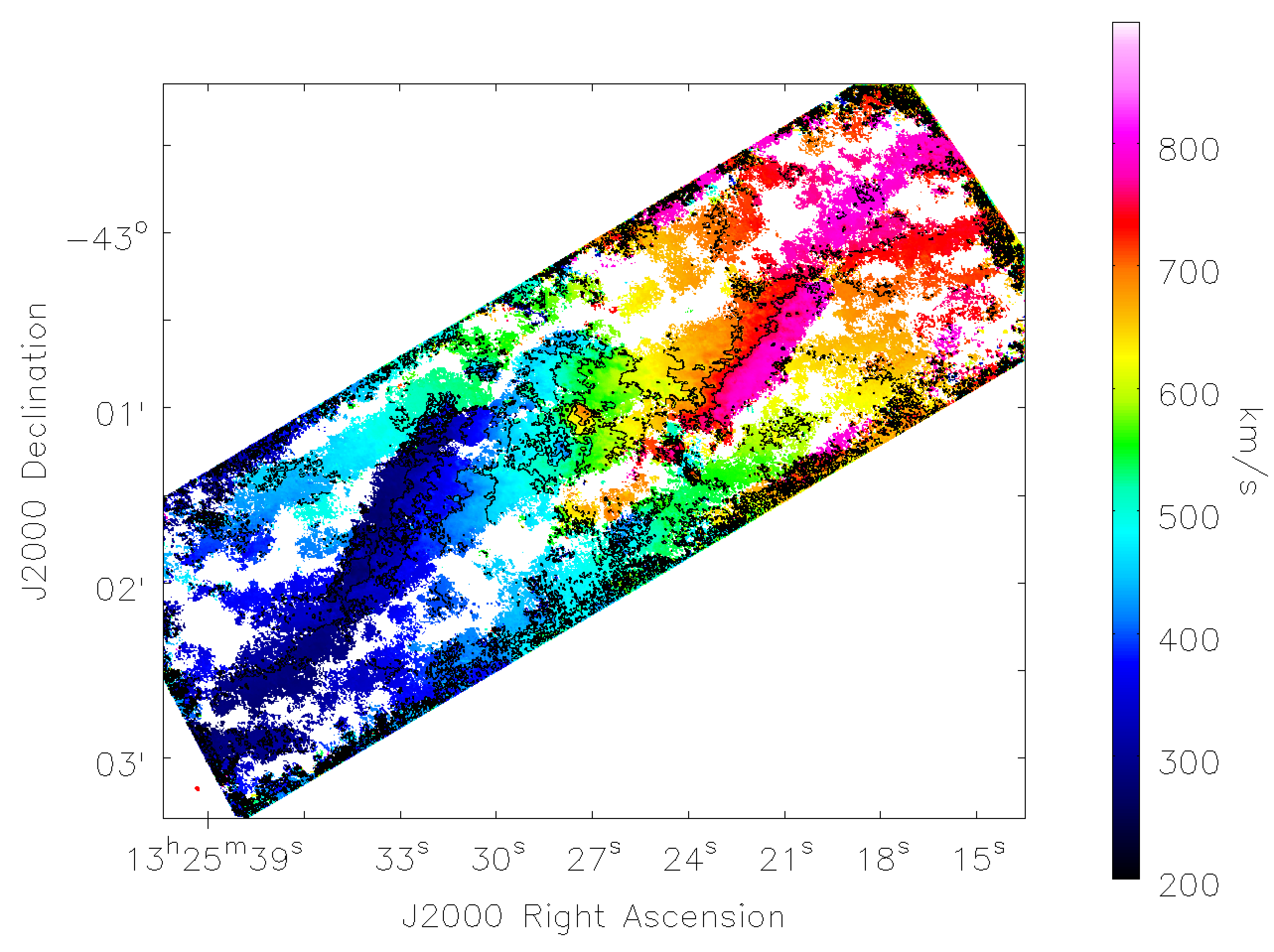}
\includegraphics[width=14cm]{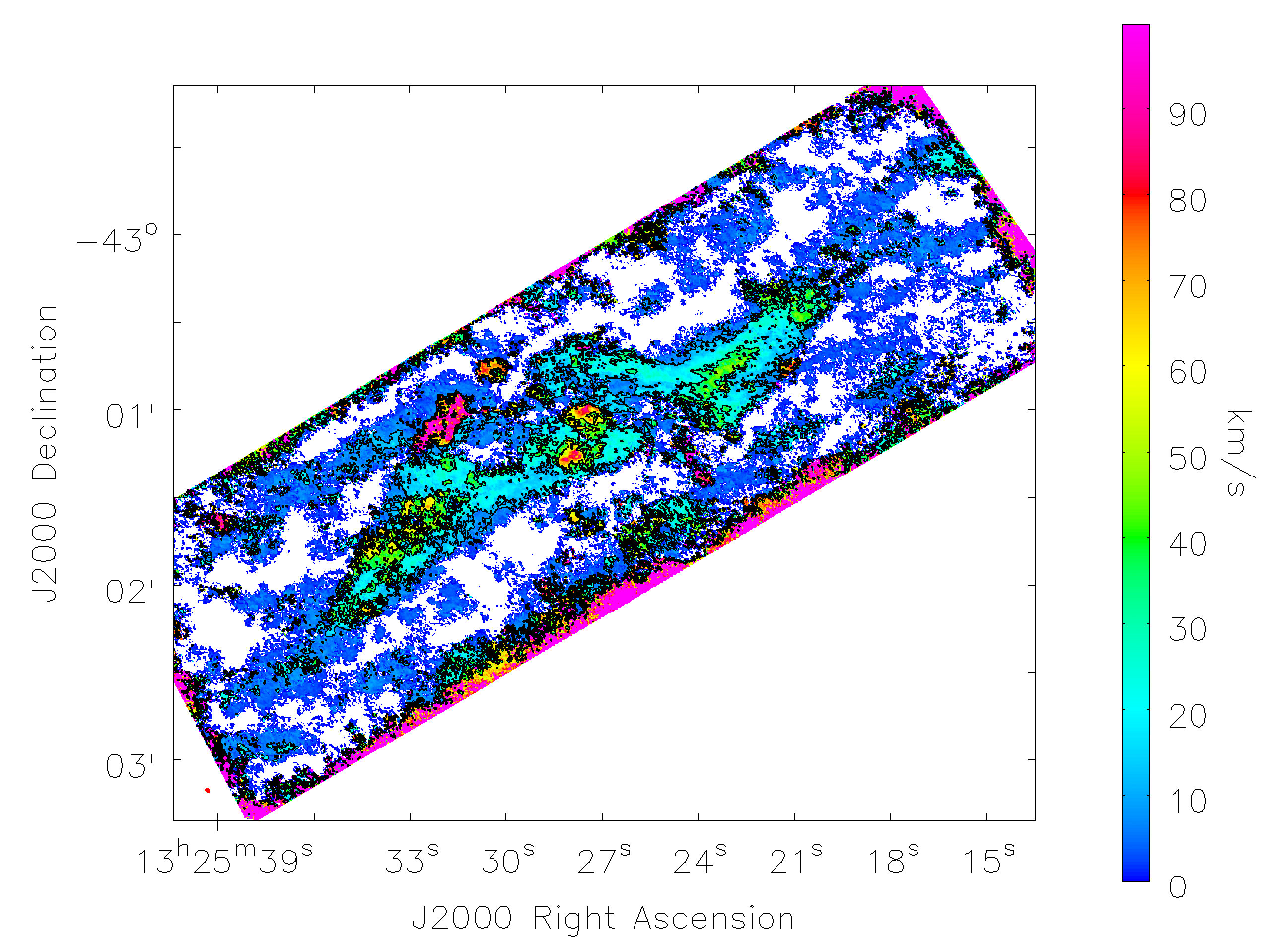}
\end{center}
\caption{
{\bf Top)} Intensity-weighted velocity field (moment 1) of the molecular disk of Centaurus~A. Contours range from 300 to 800\,\kms\, in bins of 50\,\kms. {\bf Bottom) } Velocity dispersion map (moment 2). The contour levels are 10, 30 and 50 \kms . 
\label{mom1}}
\end{figure*}

We are able to resolve for the first time the molecular component along the dust lane of Cen~A into tens of parsec scale molecular clouds thanks to ALMA's high resolution, sensitivity and high dynamic range.
 Figs.\,\ref{mom} and \ref{mom1} show the CO(1--0) integrated-intensity map, velocity field and velocity dispersion map of the molecular disk of Cen~A. 

\begin{figure*}
\begin{center}
\includegraphics[width=16cm]{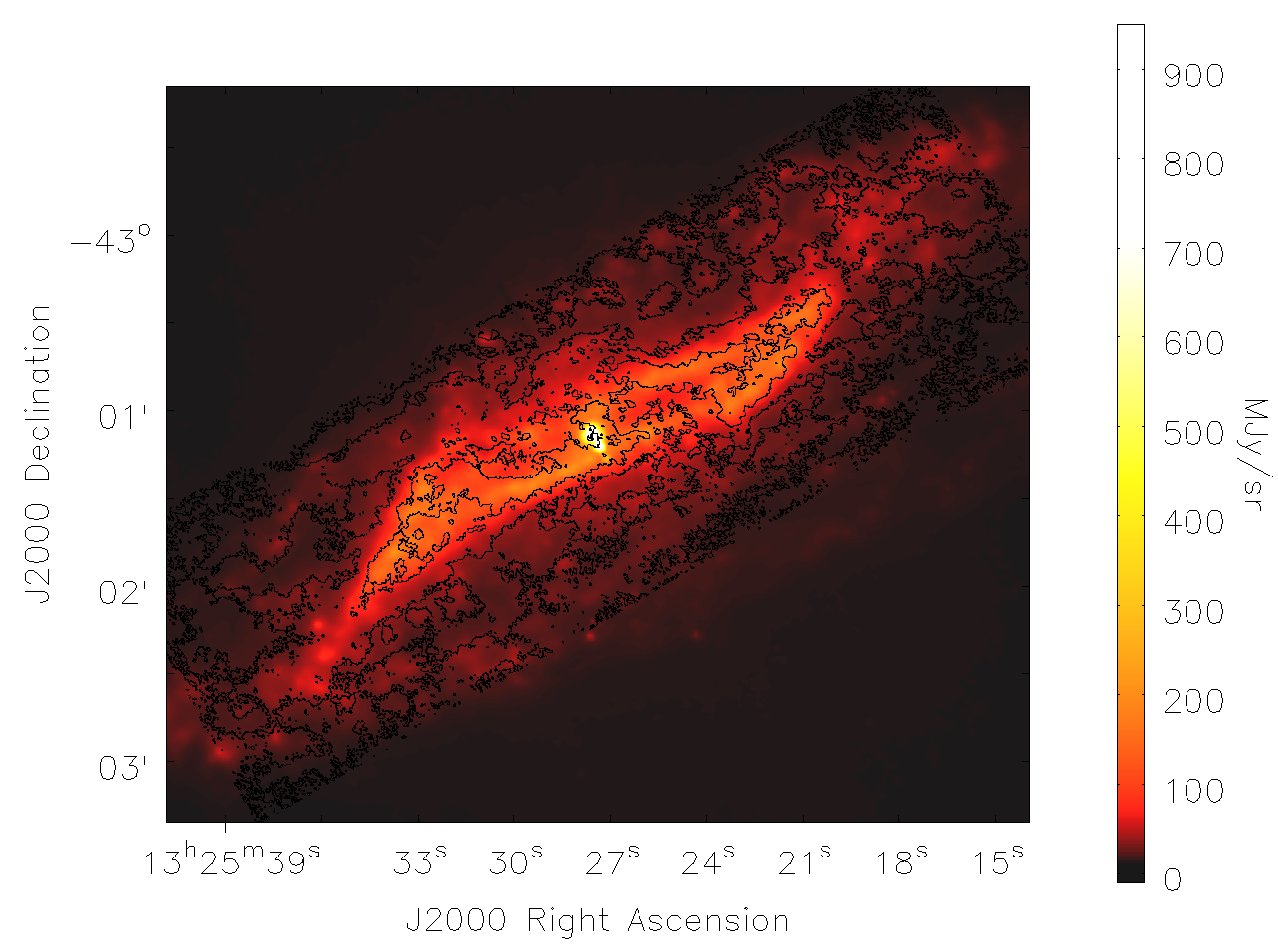}
\end{center}
\caption{
 Spitzer/IRAC 8\,$\mu$m map of Centaurus~A in color scale shown with the CO(1--0) integrated intensity contours. The contour levels are 1 and 13\,$\sigma$, where $\sigma$=0.14\,Jy\,beam$^{-1}$\kms. 
\label{8um}}
\end{figure*}

Fig.\,\ref{8um} presents the IRAC 8\,$\mu$m over the CO(1--0) contours for comparison.
The peculiar morphology seen in the 8\,$\mu$m emission (also in the 24\,$\mu$m emission) is usually referred to as the "parallelogram structure" ($\sim$3 kpc in size), which can be partially reproduced by a model of a warped and thin disk seen in projection \citep{2006ApJ...645.1092Q}. This model can match partly the distribution of CO emission, but there are some differences such as the lack of emission on the NE and SW sides of that parallelogram, as well as the curvature of CO(1-0) emission in the form of spiral arms, which are not as clear in the 8\,$\mu$m map \citep{2009ApJ...695..116E,2012ApJ...756L..10E}. It is not clear within the warped and thin disk model how the gas of the CND is maintained, since an assumption to reproduce the observed morphologies is the existence of a gap of cold gas and dust from $r$ = 0.2\,kpc to 0.8\,kpc \citep{2006ApJ...645.1092Q,2009ApJ...695..116E}.

The CO emission is distributed mainly in three distinct regions: the large scale disk, the spiral arm features, and the circumnuclear disk (CND). 
The distribution and kinematics of the CND and spiral arm regions within the inner 1\,kpc are in agreement with previous SubMillimeter Array (SMA) CO(2--1) observations \citep{2009ApJ...695..116E}. The extension of the spiral arms beyond the 1\,kpc scale, reported in \citet{2012ApJ...756L..10E} using SMA CO(2--1) mosaic  observations, is also in agreement, although their field of view was smaller, and the image was less sensitive and characterized by a poorer dynamic range than the one presented here.
Besides the main CO components previously observed, the ALMA CO(1--0) map reveals a much shallower and extended molecular disk relative to the previous SMA data, and is composed of multiple filamentary structures covering the dust lane.

{ The velocity field of Cen~A (Fig.\,\ref{mom1}, top panel) shows that the outer disk component follows the kinematics of the parallelogram disk  (i.e. receding side on the NW, and approaching on the SE). However, the different filaments in the outermost regions are kinematically distinct to those in the parallelogram structure. Other peculiar components appear at $\sim$30\arcsec\ ($\sim$500\,pc) to the N and S of the CND, which is probably gas entrained by the jet or associated with the shell-like structure reported by \citet{2006ApJ...641L..29Q}. 
The majority of the clouds in the molecular disk outskirts have velocity dispersions of about $\sim$5\,\kms\ (Fig.\,\ref{mom1} bottom), except along the arms and parallelogram structure, where dispersions are $\sim$ 20\,\kms, and are largest towards the CND. The large velocity dispersions found in the CND are compatible with the results presented in \citet{2017ApJ...843..136E} using the CO(3--2) transition, but as explained there, we note that the velocity dispersions do not exceed $\sim$40\,\kms .  Large velocity dispersions above 40\,\kms\ , in the CND or elsewhere, are usually composed of multiple lines along the line of sight. }

\subsection{Molecular Gas Mass}
\label{moleculargasmass}
 First we obtained the CO luminosities, given by:
 
 \begin{equation}
  L_{\rm CO(1-0)} = (c^2/2k_B)\,S_{\rm CO(1-0)}\,\nu_{\rm obs}^{-2}\,D_{\rm L}^2, 
 \end{equation} 
 
\noindent where $c$ is the light speed, 
$k_B$ is the Boltzmann constant, 
$S_{\rm CO}$ is the integrated CO line flux in Jy\,\kms, 
$\nu_{\rm obs}$ is the observed rest frequency in GHz, 115.271\,GHz, 
and 
$D_{\rm L}$ is the luminosity distance to the source in Mpc \citep[e.g.][]{2005ARA&A..43..677S}. 
It yields:
 \begin{equation}
  L_{\rm CO(1-0)}=2445\times S_{\rm CO(1-0)}\,D_{\rm L}^2 [{\rm K\,\kms\,pc^2}]  ,
  \end{equation}   
\noindent  and for the case of Cen~A  we obtain: 
 \begin{equation}  L_{\rm CO(1-0)} = 3.47 \times 10^8\,{\rm K\,\kms\,  pc}^{2}  .  \end{equation} The agreement of this luminosity from the combined dataset with that of the ALMA CO(1--0) single dish data assures that we recover all the flux, as shown also by the profiles in Fig.\,\ref{co1-0spectra}. We also note that single dish data are very important to achieve this. Only 47\% of the total flux in the ALMA CO(1--0) single dish data is recovered by the 7m array data.

The luminosity estimates using single dish maps from the literature based on CO(1--0) data (SEST) were found to be 1.73 $\times$ 10$^8$\,K\,\kms\,pc$^{2}$ (corrected by the different distance convention, from 3 to 3.8\,Mpc) \citep{1993A&A...270L..13R,1990ApJ...363..451E}, lower by about a factor of two than ours. This is partly caused by the larger area covered by our map and to extended emission. 
A comparison of the integrated flux in our combined map and the pointings along the parallelogram structure in \citet{2014A&A...562A..96I}, for a common region encompassing 116\arcsec\ $\times$ 45\arcsec\ along a P.A. of 125\arcdeg , agrees to within 10\% (around 4500\,Jy\,\kms ).

 We used as conversion factors between the CO integrated intensity and H$_2$ column density,  $X_{\rm CO}$
  $= N_{\rm H_2} /I _{\rm CO}$,  $X_{\rm CO}$ = 2 $\times$ 10$^{20}$~cm$^{-2}$~(K~km~s$^{-1}$)$^{-1}$ for the outer disk, and 
  $X_{\rm CO}$ = 5 $\times$ 10$^{20}$~cm$^{-2}$~(K~km~s$^{-1}$)$^{-1}$ for the CND. 
The factor for the external regions of the disk mass was obtained in Paper II using the virial method (i.e. the comparison of the CO luminosity and virial masses of identified GMCs in the Cen~A molecular disk) using the same CO(1--0) data that are presented here. This factor matches the recommended factor for the disk of the Milky Way \citep[e.g.][]{2001ApJ...547..792D} and other nearby disk galaxies \citep{2013ARA&A..51..207B}.
As for the CND,  the larger factor we assume is motivated by the value obtained by \citet{2014A&A...562A..96I} using global single-dish measurements of the CO spectral line energy distribution towards the CND and { Large Velocity Gradient (LVG)} analysis, $X_{\rm CO}$ = 4 $\times$ 10$^{20}$~cm$^{-2}$~(K~km~s$^{-1}$)$^{-1}$. In Paper II, we also confirm a larger factor using the virial method, $X_{\rm CO}$ = $(5 \pm 2) \times10^{20}$\,cm$^{-2}$(K\,\kms)$^{-1}$, in agreement within the uncertainties. In contrast to this result it should be noted that in the circumnuclear regions of starburst galaxies and mergers, as well as in the Galactic center, lower factors are usually found down to a full order of magnitude \citep[e.g.][]{2013ARA&A..51..207B}. 
Larger values than the Milky Way disk $X_{\rm CO}$ factor are usually found in low metallicity dwarf galaxies \citep[e.g.][]{2013ARA&A..51..207B,2018ApJ...864..120M}.
{ The larger factor is probably caused by high excitation conditions coupled with CO-dark H$_2$ gas, but it is not likely due to lower metallicities (Paper II). }

Finally, we calculated the molecular gas masses as in \citet{2013ARA&A..51..207B}:

\begin{equation}
 M_{\rm mol} [{\rm M}_\odot] = 4.3 \times X_{\rm 2}~L_{\rm CO(1-0)} , 
 \end{equation}
where the factor 1.36  for elements other than hydrogen \citep{2000asqu.book.....C} is taken into account, and $X_{\rm 2}$ is the $X_{\rm CO}$ factor normalized to 2 $\times$ $10^{20}$\,cm$^{-2}$\,(K\,\kms)$^{-1}$.

The total molecular gas mass in the observed field is $M_{\rm mol}$ $\simeq$ 1.6 $\times$ 10$^{9}$\,$M_\odot$. 
The smaller CO luminosity found in the literature as discussed above is reflected in a lower reported molecular gas mass (i.e. $M_{\rm mol}$ = 0.7 $\times$ $10^{9}$\,$M_\odot$, corrected by different distances, X factors and including the 1.36 factor to account for He), but they are consistent when considering the smaller area covered (i.e. the parallelogram structure). Our CO(1--0) map is sensitive to more extended gas that is beyond the parallelogram structure.
In Table~\ref{tbl3} we show the derived molecular gas masses of different regions along the disk (i.e. CND, spiral arms, outer disk, etc.).

\section{Star Formation Rate}
\label{sfr}

\subsection{Validity of using 8\,$\mu$m PAH Data as Star Formation Rate Tracer}
One of the goals of the present work is to disentangle how the SF activities may vary in the different environments of this galaxy, from the CND close to the AGN to the outermost parts of the molecular disk.
We need to reach a high spatial resolution to investigate the KS law locally in Cen~A. Approaching the excellent spatial resolution ($\sim 18$\,pc) of the ALMA maps for the SFR is not possible but the use of the 8\,$\mu$m instead of 24\,$\mu$m emission to trace the SFR mitigates this issue by a factor of 2 \citep{2017MNRAS.466.1192M}, and so we chose the 8\,$\mu$m band for our analysis because of its higher angular resolution (2\farcs4, or 43\,pc) compared to that of the 24\,$\mu$m data (6\farcs0, or 108\,pc). Moreover, \citet{2019MNRAS.483..931E} showed that the 8\,$\mu$m PAH emission is a robust SFR tracer down to physical spatial resolutions of 49\,pc.

PAH emission represents generally more than 70\% of the 8\,$\mu$m emission \citep{2011EAS....46..133C} and dominates by more than a factor of 100 above the warm dust emission and about a factor of 5 above the stellar continuum \citep{2001ApJ...554..778L, 2005ApJ...633..857D, 2014MNRAS.443.2711Y}. As a consequence, PAH emission (and in particular from the {\it Spitzer} 8\,$\mu$m band) has been used in the literature to estimate SFR in galaxies \citep{2005ApJ...633..871C, 2005ApJ...632L..79W, 2006ApJ...648..987P,2008ApJ...686..155Z,2010ApJ...723..530P, 2011EAS....46..133C, 2012ARA&A..50..531K, 2014MNRAS.443.2711Y, 2017ApJ...850...68C, 2017A&A...605A..74K, 2017MNRAS.466.1192M, 2018ApJ...865..154H,2019MNRAS.483..931E, 2019MNRAS.482..560M}. Nevertheless, there may be variations from galaxy to galaxy, due to metallicity or different beam filling factors \citep{2005ApJ...628L..29E, 2005ApJ...633..871C, 2006ApJ...648..987P, 2014MNRAS.443.2711Y}, so it is necessary to assess the validity of the 8\,$\mu$m PAH emission as a reliable star formation tracer in Cen\,A.

\begin{figure*}
\centering
\includegraphics[width=8.5cm]{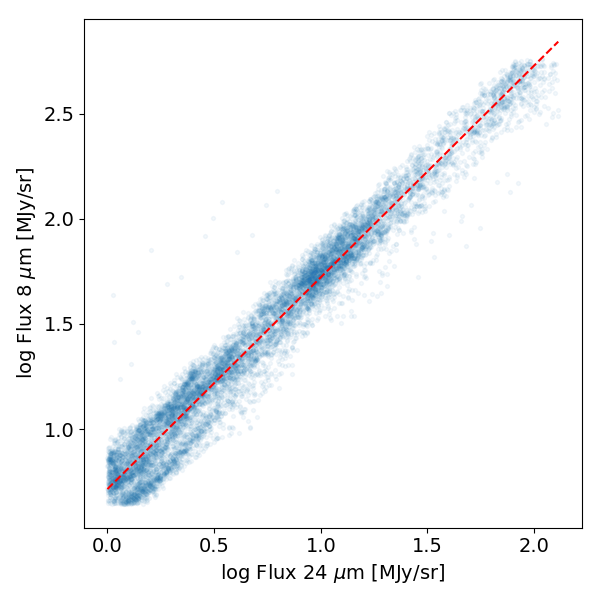}
\includegraphics[width=8.5cm]{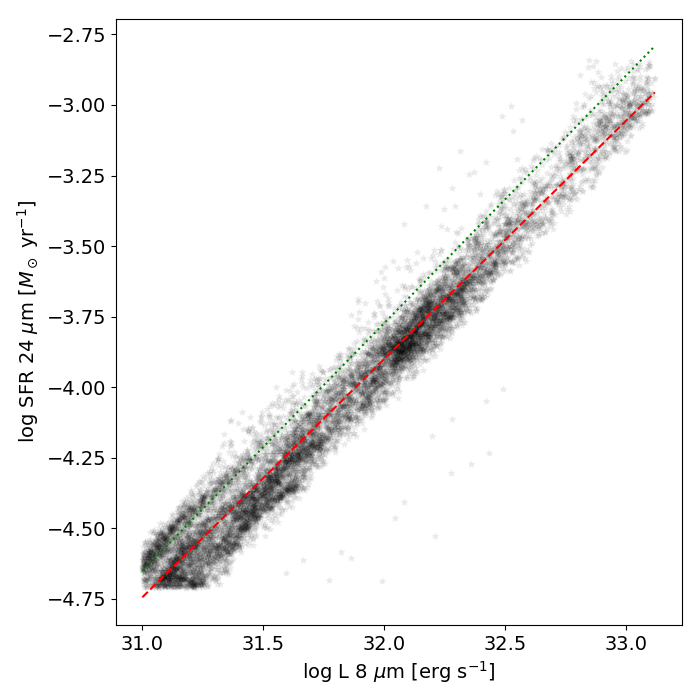}
\caption{{\bf Left)} Correlation between the 8\,$\mu$m (PAH)  and 24\,$\mu$m flux densities in decimal log scale. The (red) dashed line shows the linear fit to the data points, with a slope of 1.01 and correlation coefficient of 0.98.  {\bf Right) } Correlation between the $\Sigma_{\rm SFR}$ (obtained from 24\,$\mu$m following the calibration in \citealt{2013seg..book..419C}) and the 8\,$\mu$m PAH luminosity surface densities ($\Sigma_{L(\rm 8\,\mu{\rm m}\,PAH)}$)  in decimal log scale. 
The (red) dashed line shows the linear fit to the data points.
The calibration provided by \citet{2011AJ....142..111B} in their Table\,2 for M\,33 is also shown for comparison as a (green) dotted line. 
\label{compare24and80}}
\end{figure*}

To do so, we compared the 8\,$\mu$m PAH emission of Cen\,A to the more robust and widely used 24\,$\mu$m emission \citep{2005ApJ...633..871C, 2006ApJ...650..835A, 2006ApJ...648..987P}. 
In Fig.\,\ref{8um} we show the 8\,$\mu$m map (with CO(1--0) contours for comparison). 
We checked the spatial pixel-to-pixel agreement between 8\,$\mu$m PAH and 24\,$\mu$m fluxes ($F_{\rm 8\,\mu{\rm m}\,PAH}$ and $F_{24\,\mu{\rm m}}$, respectively, in units of MJy sr$^{-1}$) in Fig.\,\ref{compare24and80}. The 8\,$\mu$m PAH map was convolved to the FWHM of the 24\,$\mu$m map (6\arcsec). We note that in this comparison only data points outside the inner 10\arcsec\ are shown because of the difficulty to subtract the bright central AGN component in the 24\,$\mu$m map, which is worsened by the complex shape of the point spread function. Also we note that we clipped some data points at the low noise level end. We find a very tight (with a correlation coefficient of 0.98) linear correlation which can be expressed as:
\begin{equation}
\label{equation5}
\log F_\nu(\rm 8\,\mu{\rm m}\,PAH) = 1.01 \times \log F_\nu(24\,\mu{\rm m}) + 0.71 \quad .
\end{equation}

In Cen\,A, the linear coefficient against the 24\,$\mu$m emission and the very tight correlation over more than two orders of magnitude ensures that the 8\,$\mu$m PAH emission is a reliable SFR tracer  which, hence, could be linked to the hot dust reprocessing photons from star forming regions (as traced by the 24\,$\mu$m emission).
\citet{2011MNRAS.410.1197C}  found that, in a sample of 12 elliptical and lenticular galaxies, the 8\,$\mu$m PAH and the 24\,$\mu$m emissions yield similar SFR estimates.
 Therefore the 8\,$\mu$m PAH emission will be used to study the spatially resolved KS law in \S\,\ref{subsec:resolvedsflaw}. In the next subsection we will estimate the global SFR in Cen\,A from a variety of standard SFR tracers.

\subsection{Global Star Formation Rate from Various Tracers}

A large number of luminosity-to-SFR calibrations can be found in the literature and these conversions can vary by up to $\sim 50$\% and create artificial offsets between different works \citep{2012ARA&A..50..531K}, so it is necessary to use homogenized calibrations. In this section, we use the homogenized SF calibrations compiled by \citet{2013seg..book..419C} \citep[from works by][]{2005ApJ...633..871C, 2007ApJ...666..870C, 2007ApJ...671..333K, 2009ApJ...703.1672K, 2011ApJ...735...63L, 2011ApJ...741..124H}. All the calibrations make the implicit assumption that the stellar initial mass function (IMF) is constant across all environments and given by the double-power law expression with slope $-1.3$ in the range $0.1-0.5\,M_\odot$ and slope $-2.3$ in the range $0.5-120\,M_\odot$ \citep{2001MNRAS.322..231K} unless noted otherwise.

First, we use a mixed indicator combining the FUV flux from massive stars as well as its dust-obscured counterpart \citep{2013seg..book..419C}:
\begin{equation}
\begin{split}
{\rm SFR_{\rm FUV+24\,\mu{\rm m}}}\ [M_\odot\,{\rm yr}^{-1}] = \\
4.6 \cdot 10^{-44} \times [L({\rm FUV}) + 3.89 \cdot L({24\,\mu{\rm m}})] \quad ,
\end{split}
\end{equation}

\noindent  where $L$ is the luminosity in units of erg\,s$^{-1}$ for a wavelength (or range) $\lambda$. We obtained $L{\rm (FUV)}$ =  4.32 $\times$ 10$^{42}$\,erg\,s$^{-1}$ and $L{\rm (24\,\mu{\rm m})}$ = 5.19 $\times$ 10$^{42}$\,erg\,s$^{-1}$. This leads to ${\rm SFR_{\rm FUV+24\,\mu{\rm m}}} = 1.13 \pm 0.09\,M_\odot\,{\rm yr}^{-1} $ for Cen\,A.
Second, we use the monochromatic calibration by \citet{2009ApJ...692..556R} referenced to the same \citet{2001MNRAS.322..231K} IMF to be consistent with the other calibrations by \citet{2013seg..book..419C}:

\begin{equation}
{\rm SFR}_{24\,\mu{\rm m}}\ [M_\odot\,{\rm yr}^{-1}] = 2.04 \times 10^{-43} L({24\,\mu{\rm m}})\ [{\rm erg\,s}^{-1}] \quad ,
\end{equation}

\noindent and it yields ${\rm SFR}_{24\,\mu{\rm m}} = 1.06 \pm 0.11\,M_\odot\,{\rm yr}^{-1} $.
Another standard monochromatic calibration, based on the 70\,$\mu$m emission \citep[Eq.~22 in][]{2010ApJ...714.1256C}, reads:

\begin{equation}
{\rm SFR}_{70\,\mu{\rm m}}\ [M_\odot\,{\rm yr}^{-1}] = 5.9 \times 10^{-44} L({70\,\mu{\rm m}})\ [{\rm erg\,s}^{-1}] \quad .
\end{equation}

\noindent   We obtain $L{\rm (70\,\mu{\rm m})}$ = 1.95 $\times$ 10$^{43}$\,erg\,s$^{-1}$ (from the 70\,$\mu$m flux in \citealt{2012MNRAS.423..197B}), and it yields $ {\rm SFR}_{70\,\mu{\rm m}} = 1.15 \pm 0.23\,M_\odot\,{\rm yr}^{-1} $.
Finally, we estimate the total IR (TIR) flux (in W\,m$^{-2}$) in the range $3-1000\,\mu$m,  $F({\rm TIR})$, by using the following relation \citep{2002ApJ...576..159D,2010A&A...510A..64V}:

\begin{equation}
\begin{split}
 F({\rm TIR}) = 10^{-14} \times [19.5\ F_\nu(24\,\mu{\rm m}) + 3.3\ F_\nu(70\,\mu{\rm m}) \\
 + 2.6\ F_\nu(160\,\mu{\rm m})] \quad ,
 \end{split}
\end{equation}

\noindent where $F_\nu(24\,\mu{\rm m})$, $F_\nu(70\,\mu{\rm m})$, and $F_\nu(160\,\mu{\rm m})$ are the MIPS flux densities in Jy for Cen\,A as listed in Table 3 of \citet[][]{2012MNRAS.423..197B}. This TIR flux turns into a luminosity of $L(\rm TIR) = 1.29 \pm 0.16 \times 10^{10}\,L_\odot$. We can then derive the SFR based on dust-processed stellar light from the TIR luminosity using the following calibration \citep{2013seg..book..419C}, where the assumptions are a mass range of the stars in the IMF spanning $0.1-100\,M_\odot$ and the SF remaining constant during $100$\,Myr:

\begin{equation}
{\rm SFR_{TIR}}\ [M_\odot\,{\rm yr}^{-1}] = 2.8 \times 10^{-44}\, L({\rm TIR})\ [{\rm erg\,s^{-1}}] \quad ,
\end{equation}

\noindent which leads to SFR$_{\rm TIR}$ = 1.38 $\pm$ 0.17\,$M_\odot$\,yr$^{-1}$ for Cen\,A.
All these estimates of global SFR lead to a consistent result for a SFR in Cen\,A of about 1\,$M_\odot$\,yr$^{-1}$. 

The SFRs we have obtained are also in agreement with previous measurements in the literature obtained from IRAS data. The total infrared luminosity was estimated as $L({\rm TIR})$ $\approx$ 2 $L({\rm FIR}) = 1.12 \times 10^{10}\,L_\odot$  \citep{2006A&A...447...71V,1988A&A...193...27M} (corrected to our distance convention distance). Without the contribution of the AGN as estimated by \citet{2006A&A...447...71V}, then $L({\rm TIR}) \approx 9.4 \times 10^9\,L_\odot$. The total SFR is estimated to be $\approx 1\,M_\odot\,{\rm yr}^{-1}$ \citep{2004ApJ...602..231C,2006A&A...447...71V}. 
\citet{2004ApJ...602..231C} estimated SFR = $1\,M_\odot\,{\rm yr}^{-1}$ (converted to our distance convention) from a combination of $L({\rm FIR})$ and UV luminosity, $L({\rm UV})$, and a conversion factor to convert from $L({\rm FIR+UV})$ to SFR of $5.7 \times 10^{-44}$. However, the contribution of the AGN was not subtracted.  
\citet{2006A&A...447...71V} obtained SFR = $1.6\,M_\odot\,{\rm yr}^{-1}$ (also converted to our distance convention) from $L({\rm TIR})$ and after subtraction of the AGN contribution. 
In this estimate, \citet{2006A&A...447...71V} used the SFR calibration of \citet{1998ARA&A..36..189K}, SFR$_{\rm TIR}$ ($M_\odot\,{\rm yr}^{-1}) = 4.5 \times 10^{-44} L({\rm TIR}) ({\rm erg\,s}^{-1}$). To be able to compare this with our analysis we must convert it to a Kroupa IMF, and therefore we divided it by a factor of 1.59 \citep{2008AJ....136.2846B}. Finally, we obtain SFR $\simeq$ 1\,M$_\odot$~yr$^{-1}$, which is consistent with all the previously mentioned estimates.

 As the 8\,$\mu$m PAH emission is so closely tied to the 24\,$\mu$m emission (see Fig.\,\ref{compare24and80}), we can provide a calibration of the global SFR in Cen\,A from the 8\,$\mu$m PAH emission to reproduce the global SFR obtained from the 24\,$\mu$m emission (i.e. $1.06\,M_\odot\,{\rm yr}^{-1}$):
\begin{equation}
\begin{split}
{\rm SFR}_{\rm 8\mu m\,PAH}\ [M_\odot\,{\rm yr}^{-1}] = \\ 
6.14 \times 10^{-44} L({\rm 8\mu m\,PAH})\ [{\rm erg\,s}^{-1}] \quad .
\end{split}
\end{equation}

\section{Spatially Resolved Kennicutt-Schmidt Star Formation Law}
\label{subsec:resolvedsflaw}

We present the spatially resolved KS SF law using the ALMA CO(1--0) data to derive the molecular gas surface densities ($\Sigma_{\rm mol}$), along with the Spitzer 8\,$\mu$m PAH emission. 
We used a common astrometric grid and we aligned both maps so that there are no artificial offsets. The pixel size in our analysis is 2\farcs4 (or 43\,pc, i.e. that of the 8\,$\mu$m map).

We derived the molecular surface densities, $\Sigma_{\rm mol}$, from the CO(1--0) map presented in \S\,\ref{co1-0} and following the convention explained in \S\,\ref{moleculargasmass}.
The map is in units of $M_\odot$\,pc$^{-2}$ and includes the contribution of helium and other heavy elements (a factor 1.36).
We estimate that our $\Sigma_{\rm mol}$ map is complete down to surface densities of  $\Sigma_{\rm mol} \simeq 4\,M_\odot\,{\rm pc}^{-2}$ (corresponding to 3$\sigma$ and a velocity width of 10\,\kms).

To obtain the SFR surface densities, $\Sigma_{\rm SFR}$, from 8\,$\mu$m PAH emission (old-stellar contribution corrected as explained in \S\,\ref{spitzerdata}), we take advantage of the very tight correlation between the 24\,$\mu$m and the 8\,$\mu$m PAH emissions. We estimate the SFR from the 24\,$\mu$m emission using the local recipe by \citet{2007ApJ...666..870C} calibrated on 220 H{\sc ii} knots in star forming regions from 33 nearby galaxies. A linear fit to recover the local SFR from 8\,$\mu$m PAH emission leads to:
\begin{equation}
\log(\Sigma_{\rm SFR}) = 0.844 \times \log(\Sigma_{L(\rm 8\,\mu{\rm m}\,PAH)}) - 30.923 \quad ,
\end{equation}
where $\Sigma_{\rm SFR}$ is in units of $M_\odot\,{\rm yr}^{-1}\,{\rm kpc}^{-2}$ and $\Sigma_{L(\rm 8\,\mu{\rm m}\,PAH)}$ is the average 8\,$\mu$m PAH luminosity surface density in ${\rm erg\,s}^{-1}\,{\rm kpc}^{-2}$.

 Similarly, other authors have already reported that the 8\,$\mu{\rm m}$ PAH emission 
correlates with the SFR, but follows a power law with a coefficient 
slightly less than unity. The sub-linear behaviour has been reported by 
\citet{2005ApJ...633..871C}, \citet{2006ApJ...648..987P}, \citet{2014MNRAS.443.2711Y}, \citet{2017ApJ...850...68C} and \citet{2019ApJ...873....3T}, among others. { The sub-linear trend is similar to that usually found for the 24\,$\mu$m emission \citep[see for example][]{2007ApJ...666..870C,2011AJ....142..111B}, as expected because of the strong linear correlation we obtained between the 8\,$\mu$m PAH and 24\,$\mu$m emissions (Eq.\,\ref{equation5}).}
In Fig.\,\ref{compare24and80} (right) we show the result of the linear fit as well as the calibration provided by \citet{2011AJ....142..111B} in their Table\,2. The two fits have a very similar slope but the \citet{2011AJ....142..111B} intercept is slightly higher than the one we find. The recipe provided by \citet{2011AJ....142..111B} has been calibrated in the M\,33 galaxy, where a relatively high conversion factor is indicative of a rather low dust content \citep{2009A&A...493..453V}.

Our surface density maps have been corrected by a fixed inclination of 70$^\circ$. This is a good approximation for the average inclination of the disk. However, we note that the disk is likely warped, and the inclination may range from 60 to 120$^\circ$ \citep{2010PASA...27..396Q}. Fortunately, the uncertain inclination for each pixel and the possible issue of having different components along the line of sight should affect equally both $\Sigma_{\rm mol}$ and $\Sigma_{\rm SFR}$.

\subsection{Kennicutt-Schmidt Star Formation Law}

\begin{figure*}[htbp]
\begin{center}
\includegraphics[width=12cm]{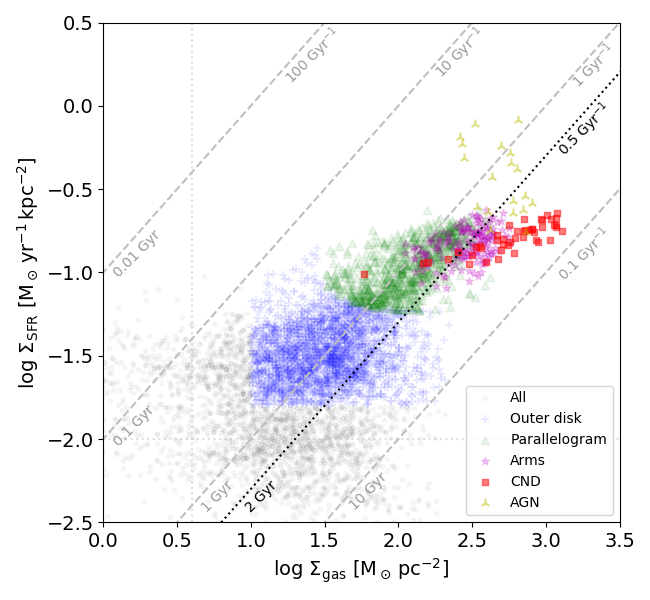}
\includegraphics[width=15cm]{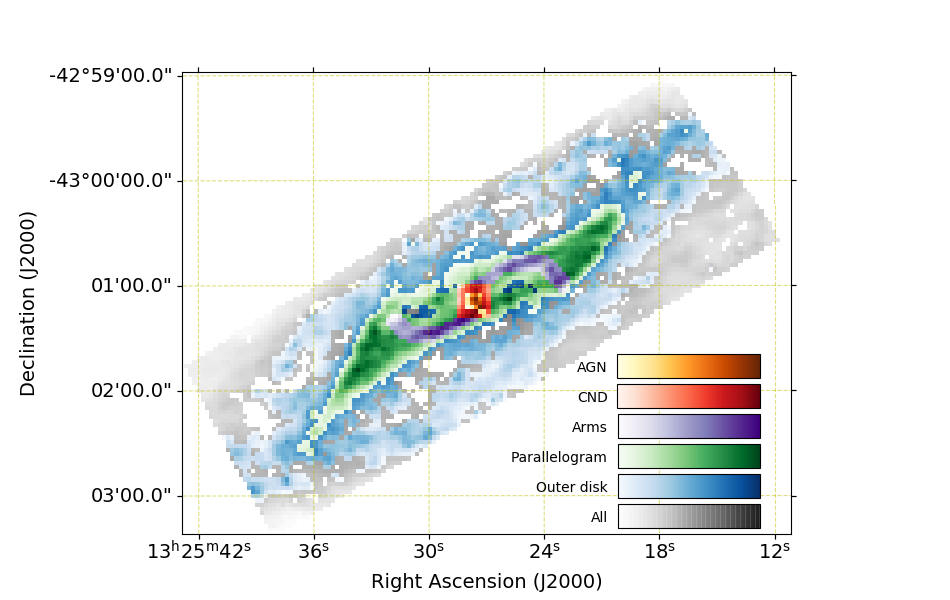}
\end{center}
\caption{ 
{\bf Top)} The Kennicutt-Schmidt star formation law plot of the molecular disk of Centaurus~A: star formation rate surface density ($\Sigma_{\rm SFR}$), estimated from 8\,$\mu$m PAH emission, as a function of molecular gas surface density ($\Sigma_{\rm mol}$) from the CO(1--0) emission. Each data point corresponds to a 2\farcs4 pixel (resolution of the 8\,$\mu$m data).  Dashed lines indicate constant SFEs at 0.1, 1, 10, and 100\,Gyr$^{-1}$ (or depletion times 0.01, 0.1, 1, and 10\,Gyr). Dotted (vertical and horizontal) lines show the estimated sensitivity limits for $\Sigma_{\rm mol}$ and $\Sigma_{\rm SFR}$. The dotted line at constant SFE 0.5\,Gyr$^{-1}$ (or depletion time 2\,Gyr) is displayed as reference for the SF law of disk galaxies \citep[e.g.][]{2008AJ....136.2846B}.
The CND \citep{2009ApJ...695..116E} is marked as red squares, the spiral arms \citep{2012ApJ...756L..10E} as purple star symbols, the parallelogram structure \citep{2006ApJ...645.1092Q} as green triangles, and the outer disk as blue crosses. 
Other regions are shown in grey color, and data-points that might be contaminated by the AGN are shown as yellow triangle symbols. {\bf Bottom)}
Scheme showing the main regions color coded as in the upper panel.
}
\label{figsfr}
\end{figure*}

We present in Fig.\,\ref{figsfr} the spatially resolved KS SF law plot (i.e. $\Sigma_{\rm SFR}$ versus $\Sigma_{\rm mol}$) along the dust lane of Centaurus~A. The scatter is large, but by visual inspection 
it is clear that there are two main clusters of data points, one cluster at high $\Sigma_{\rm mol}$ and $\Sigma_{\rm SFR}$, and another centered at $\Sigma_{\rm mol}$ and $\Sigma_{\rm SFR}$ about 1\,dex smaller.

We used a machine learning technique to separate the dataset in the $\log \Sigma_{\rm SFR} - \log \Sigma_{\rm mol}$ parameter space (see Fig.\,\ref{figsfr}) into clusters. We first remove the 19 points corresponding to the AGN to focus only on separating the low and high components in this parameter space. We use a Gaussian mixture model (GMM), based on an iterative expectation - maximization algorithm, to extract a mixture of multi-dimensional Gaussian probability distributions that best model our dataset \citep{vanderplas2016python}. To separate our dataset into two components, random initializations are used, and we tested several covariance options (diagonal, spherical, tied, and full covariance matrices). A Bayesian Information Criterion (BIC) selects the best model, which is the one obtained with the full covariance matrix, allowing each cluster to be modeled by an ellipse with arbitrary orientation.

The results of the separation are shown in Fig.\,\ref{figsfr}. This machine learning segregation is able to effectively separate the inner from the outer parts of the galaxy. The highest $\Sigma_{\rm SFR}$ -- $\Sigma_{\rm mol}$ cluster corresponds to the CND, arms, and parallelogram regions while the lowest cluster defines the outer disk region. We also use a color/symbol scheme in Fig.\,\ref{figsfr} (bottom) to separate the data points from the different regions (i.e. CND, arms, parallelogram, and outer disk). Although the separation has a physical meaning regarding regions with high or low SFR/gas surface densities in the CND, arms and outer disk, note that in the parallelogram structure regions the surface densities may artificially be higher because of the projection effect. 

{ The slope of the orthogonal distance regression fit to the data points characterizing the SF law is $N$ = 0.63 (intercept is --2.4). We note however that the scatter is large, there is a trend of decreasing slope for regions at inner radii ( for the CND $N$ = 0.28, and intercept --1.6), and also there might be further uncertainties in the fits due to projection effects, so this result should be taken with caution.}

\subsection{Star Formation Efficiency}
\label{sfe}
In this section we study the Star Formation Efficiency (SFE), defined as the SFR per unit of molecular gas mass,  $\Sigma_{\rm SFR}$ / $\Sigma_{\rm mol}$. 
The depletion time is then defined as the inverse of SFE, $\tau$ = 1/SFE.
We find a total molecular gas mass of 1.6 $\times$ 10$^9$\,$M_\odot$ and a global SFR of $\sim$1\,\Msol\ using different recipes, so the SFE is 0.6\,Gyr$^{-1}$ (or depletion time $\tau =$ 1.5\,Gyr), similar to that in spiral galaxies.
For reference in the resolved KS plot in Fig.\,\ref{figsfr} (upper panel), we indicate constant SFE lines, at SFEs 0.1, 1, 10, and 100\,Gyr$^{-1}$ (or equivalently depletion times $\tau =$ 0.01, 0.1, 1, and 10\,Gyr). 

\begin{figure*}[htbp]
\begin{center}
\includegraphics[width=16cm]{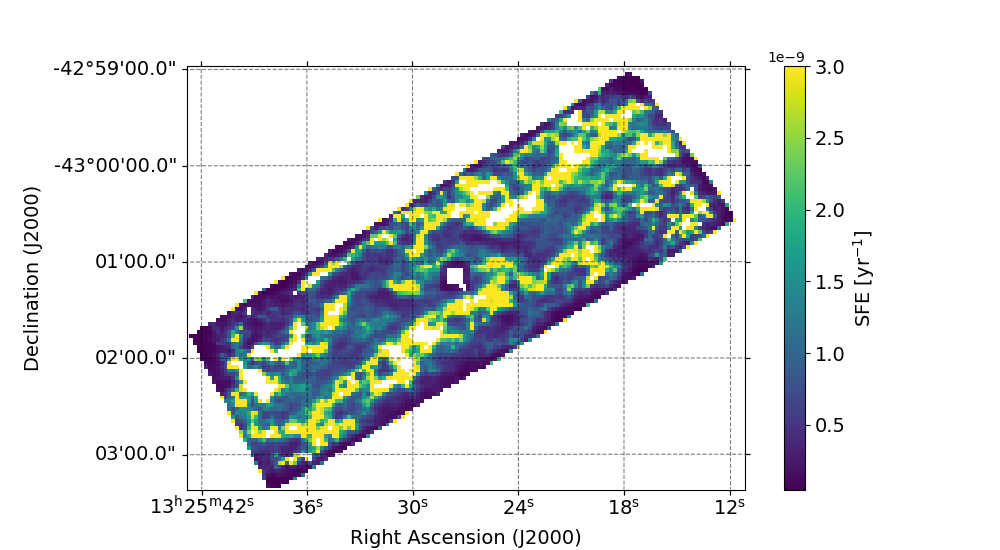}
\caption{
Pixel-to-pixel map of the star formation efficiency (SFE = $\Sigma_{\rm SFR}$ / $\Sigma_{\rm mol}$) along the dust lane of Cen~A. 
} 
\label{figsfe}
\end{center}
\end{figure*}

Fig.\,\ref{figsfe} displays the map of the SFE. 
The SFEs across the disk vary by at least two orders of magnitude from pixel to pixel. We confirm with this plot that the SFEs along the arms and the CND are lower by a factor of 2 to 4 compared to the remaining area, and that higher SFEs are found preferentially in the outskirts of the disk (blue data points in Fig.\,\ref{figsfr} bottom panel).

In columns 3 to 5 of Table~\ref{tbl3} we show  the mean, median, and standard deviation of the SFEs obtained from the pixels within the different regions. While the mean SFE in the CND and arms is 0.3 -- 0.5\,Gyr$^{-1}$, in the outer disk it is 1.3\,Gyr$^{-1}$. Also the standard deviation of the SFEs increases with radius, which likely implies variable physical conditions of molecular gas and star formation activities in the outskirts. 
This trend is contrary to what is observed in the central regions of other AGNs, as will be discussed in \S\,\ref{subsec:62}.

\section{Discussion}
\label{discussion}

\subsection{The Fate of Recently Accreted Gas in a Giant Elliptical Galaxy}

Cen~A is a rare case of a giant elliptical galaxy with a large amount of molecular gas recently accreted and with ongoing SF.
In agreement with \citet{2009AJ....137.3053Y} for other early type galaxies, the bulk of its mid-IR emission can be mostly traced to SF.
However, the SF law seems remarkably different to the trends observed for other early type galaxies. 
Thanks to its recent merger with a gas-rich galaxy,  the recent SF history varied substantially compared to other objects of its class.
The global SFE ($\sim$ 0.6\,Gyr$^{-1}$) is compatible within the uncertainties to the average of local star forming galaxies, SFE $\sim$ 1\,Gyr$^{-1}$ \citep[e.g.][]{2018ApJ...853..179T} found across different galaxy environments \citep{2011A&A...534A.102L,2012A&A...540A..96M}.

In the literature it is usually argued that SFEs are either equal or lower in early-type galaxies than the standard relation for late-type galaxies.
 The derived SFRs and molecular gas surface densities (measured globally) place E/S0 galaxies overlapping the range spanned by disk galaxies although with a larger scatter \citep{2010ApJ...725L..62W,2010MNRAS.402.2140S}. \citet{2011MNRAS.410.1197C} find that although 12 E/S0s in the SAURON project lie closely around a constant SFE, as in spiral galaxies, there are some hints of lower SFEs. 
 In fact, it was reported that molecular gas-rich early-type galaxies in the ATLAS3D project  have a median SFE of $\sim$0.4 -- 0.09 \,Gyr$^{-1}$  \citep[e.g.][]{2014MNRAS.444.3427D,2015MNRAS.449.3503D}, considerably lower than the typical SFEs of $\sim$1\,Gyr$^{-1}$ in star forming galaxies.
It has been suggested that the SF in early types is suppressed due to the stability of the molecular disks against gravitational fragmentation using the Toomre Q test \citep{2014MNRAS.444.3427D,2017ApJ...845..170B}. This result is supported by hydrodynamic models of \citet{2013MNRAS.432.1914M}, where stars form two to five times less efficiently in early-type galaxies than in spirals. \citet{2015MNRAS.449.3503D} argue that dynamical effects induced by the minor merger might be responsible for the low SFE values.
Recent studies using ALMA observations of some of the most extreme cases  in terms of low SFE bulge-dominated galaxies with dust lanes confirmed that SFE ranges from 0.002 -- 0.04\,Gyr$^{-1}$  \citep{2018MNRAS.476..122V}.
However, \citet{2017A&A...605A..74K} find that most of the ATLAS3D local early type galaxies follow the same SF law as local later type star-forming galaxies. 
Contrary to recent results suggesting slightly lower SFEs, they claim that this is directly related to the somewhat lower SFRs derived in those works. 
They note however that there is evidence that early type galaxies whose cold gas has an external origin (kinematically misaligned gas with respect to stars) have more varied SFEs, which may suggest a more bursty and variable recent SF history. This might be the case for Cen~A.

A parameter that is crucial to understand the different claims regarding the SFE in early-type galaxies is the stellar mass. In Cen~A the stellar mass is $M_\star$ = 10$^{11}$\,$M_\odot$ \citep{2013ARA&A..51..511K}. This value is within the range of the early-type galaxy sample with recent minor mergers studied by \citet{2015MNRAS.449.3503D}.
About 80\% of their low mass E/S0s ($M_\star$ $<$ 4 $\times$ 10$^{10}$\,$M_\odot$)  present high SFEs (above the KS law), and are usually star forming (i.e. blue sequence) objects \citep{2010ApJ...725L..62W}.  The fraction of E/S0s that are star forming increases with decreasing stellar mass \citep{2009AJ....138..579K}. Indeed, Cen~A's stellar mass is close to the "shutdown mass" (i.e. mass at which blue-sequence E/S0s first emerge, and above which nearly all galaxies are generally old and red) for blue sequence early type objects, $M_{\rm s}$ $\sim$ 1 -- 2 $\times$ 10$^{11}$\,$M_\odot$, which only a minority (2\%) of E/S0s exceed \citep{2009AJ....138..579K}. So in summary Cen~A is at the massive end where it is rare to find a star-forming elliptical. It is therefore likely that it is mostly by accretion events that such an object can exist.

Next we put the case of Cen~A in the context of previous works where the SFEs were found to be either equal or lower than the standard values of spiral galaxies.
Because Cen~A recently accreted material, it would match the selection criteria of galaxies in the samples of \citet{2015MNRAS.449.3503D} and \citet{2018MNRAS.476..122V}. Indeed they focus on a sample of bulge-dominated galaxies with large dust lanes and with signs of a recent minor merger. 
 In these works it is argued that the SFE is suppressed as a result of dynamical effects. How can we reconcile the apparent contradiction of Cen~A presenting a nearly standard SFE similar to star forming galaxies? This can be due to different impact parameters of the merger, merged galaxy properties (such as mass and gas fraction), and the dynamical relaxation time. 
At any rate, the case of Cen~A shows that the SFE is not always suppressed in early-type galaxies after gas-rich minor mergers. 
ALMA observations of the elliptical and radio jet galaxy NGC\,3557 have recently shown that its molecular disk within the inner hundreds of parsec presents a global SFE that may be compatible with the standard relation in spiral star-forming galaxies too \citep{2019ApJ...870...39V}.

In about 1\,Gyr, objects like Cen~A may look more similar to other objects of its class (i.e. giant elliptical galaxies without much molecular gas and with low SFEs).
A fast evolutionary sequence may be playing a role, which would explain why only a minority of massive early type galaxies are in the blue sequence even though they are in high galaxy density environments where accretion events are relatively frequent. The molecular gas will be exhausted preferentially in the external regions. 
The SFE is higher in the outskirts by a factor of four than in the central regions of Cen~A so it will be consumed faster. In addition, the gas will be either driven to the central regions, or destroyed because of a lack of self-shielding to massive SF combined with the hot environmental conditions within the elliptical galaxy (i.e. a hot ISM thermally radiating at X-rays, \citealt{1997SSRv...81....1H}).
Dense gas that is stable against collapse due to inflowing motions, shear, and/or excess of shocks/turbulence will remain, preferentially in compact components or filamentary structures within the stellar component, which is the result observed in other massive elliptical galaxies.

\subsection{The Low SFE in the CND of Cen~A: A Comparison with Low Luminosity AGNs and Nuclear Starbursts}
\label{subsec:62}

For the amount of molecular gas in the CND, the SFR is comparatively smaller than that found in other regions. The mean SFE in the CND is just about $\sim 0.3$\,Gyr$^{-1}$ at 40\,pc scales, four times smaller than in the outer regions (see \S\,\ref{sfe}). 
Using { photon-dominated region (PDR)} modeling of the neutral and ionized gas from the far-infrared lines, \citet{2017A&A...599A..53I} claimed that { because the radiation field in the CND is weaker by an order of magnitude than in the extended disk, no significant SF in the CND is required.  The radiation field strength could be explained just by the radiation coming from the stellar component of the extended disk.}
We do find that the SFR in the CND is small but not negligible. We note that this SFR is a lower limit  because we have excluded pixels close to the AGN that may be contaminated by it, but the SFE we quote is representative for the outer ($r$ =  100 -- 200\,pc) regions of the CND.

A source of uncertainty in the deduced low SFE of the CND is the $X_{\rm CO}$ factor used to calculate the molecular gas mass. For the CND 
we used a conversion factor $X_{\rm CO}$ = $5\times10^{20}$\,cm$^{-2}$(K\,\kms)$^{-1}$, while for the rest of the disk we applied $2\times10^{20}$\,cm$^{-2}$(K\,\kms)$^{-1}$. However, a different $X_{\rm CO}$ in the central regions is supported by an analysis in \citet{2014A&A...562A..96I} using LVG modeling on CO spectral line energy distribution, and Paper II using the virial method. These are independent methods so this result is likely robust. Moreover, in Paper II we found a gradual increase of $X_{\rm CO}$ towards inner radii. We note that $X_{\rm CO}$ is driven by metallicity in the low metallicity regime, but in the case of Cen~A metallicity across the molecular disk (including the CND) is $\sim 0.75$ $Z_\odot$ and relatively constant. 

The SFEs are also relatively constant at 1 -- 2\,Gyr$^{-1}$ along galactocentric radii on kpc scales \citep{2008AJ....136.2782L,2013AJ....146...19L,2019PASJ..tmp...27M}.
However, \citet{2013AJ....146...19L} systematically found higher SFEs for a given $X_{\rm CO}$ in the inner kpc of a sample of
30 nearby disk galaxies that show nuclear molecular gas concentrations, including AGN and nuclear starbursts, but not galaxy-wide starbursts.
Our finding of lower SFEs in the CND compared to the outer regions of Cen~A contradicts this.
Since  $X_{\rm CO}$ is typically smaller by a factor of 2 below the galaxy mean on average close to their nuclei \citep{2013ApJ...777....5S}, and in some cases 5 --10 times below the standard galaxy disk value, the resulting SFEs are higher than when assuming a constant $X_{\rm CO}$.

A direct comparison of SFEs between different objects requires the use of similar spatial scales. 
Using ALMA CO(2--1) maps of 14 disk galaxies, \citet{2018ApJ...861L..18U} found that on 120\,pc scales  SFEs range from 0.43 -- 1.65\,Gyr$^{-1}$, similar to the range we see for the different regions in Cen~A, from SFE = 0.3 -- 0.5\,Gyr$^{-1}$ in the CND/arms region, to SFE = 1.3\,Gyr$^{-1}$ in the outer disk.
\citet{2015A&A...577A.135C} studied the central regions (from 20 to 200\,pc) of a sample of four galaxies possessing low luminosity AGNs, i.e. Seyfert and LINER nuclei, and they found that they were characterized by higher SFEs than the standard value for disk galaxies (adopting a common $X_{\rm CO}$ factor $2.2\times10^{20}$\,cm$^{-2}$(K\,\kms)$^{-1}$), with a behaviour that is in between galaxy-wide starburst systems and more quiescent galaxies. The SFR and gas surface densities of the central regions of these objects are comparable to those in the CND of Cen\,A, ranging from $\sim$ 0.01 to 1\,$M_\odot$\,yr$^{-1}$\,kpc$^{-2}$ and $\sim$ 10$^2$ to 10$^3$\,$M_\odot$\,pc$^{-2}$, respectively. Possible causes for the lower SFEs in the CND of Cen~A are stronger shear motions and shocks. Also, the more continuous period of AGN feedback in the case of Cen~A than in other AGNs may be responsible for the observed differences. However, we also note that other late-type AGN galaxies with low SFEs towards the circumnuclear regions also exist, such as in the inner spiral regions of NGC\,1068 \citep{2012ApJ...746..129T}.  In our Galactic centre the current SFR per unit mass of dense gas was also found to be an order of magnitude smaller than in the disk \citep{2013MNRAS.429..987L}.

\section{Summary}

High angular/spectral resolution, sensitivity, and dynamic range ALMA CO(1--0) data is presented for the nearest giant elliptical and radio galaxy, Centaurus~A (NGC\,5128). 
We combined ALMA 12m, 7m and TP array data in order to recover the flux at all spatial scales. 
We performed a Kennicutt-Schmidt Star Formation law analysis at GMC scales ($\sim$ 40\,pc) using ALMA CO(1--0) data for the molecular gas and archival Spitzer mid-IR data as a tracer of the SFR. We exclude from our analysis regions where the star formation rate (SFR) tracer is contaminated by emission from the active galactic nucleus (AGN). Our main results can be summarized as follows:

\begin{enumerate}

\item  We revealed several filamentary molecular components beyond the previously known main molecular components (circumnuclear disk -CND-, arms, parallelogram structure; \citealt{2009ApJ...695..116E,2012ApJ...756L..10E}) in the outer regions of the molecular disk, extending up to a radius of $\sim$ 3\,kpc. There is a good correlation with the mid-IR continuum emission.

\item We obtained a total molecular gas mass across the dust lane of Cen~A based on CO(1--0) emission of 1.6 $\times$ 10$^9$\,$M_\odot$. This is larger than previously reported because of a larger field of view and better sensitivity to extended flux in our observations. We use a conversion factor $X_{\rm CO}$ = $2\times10^{20}$\,cm$^{-2}$(K\,\kms)$^{-1}$, except for the CND where we use $5\times10^{20}$\,cm$^{-2}$(K\,\kms)$^{-1}$ as recommended in Miura et al. (2019, Paper II) using the virial method. This is compatible with \citet{2014A&A...562A..96I} using multi-transition analysis and LVG (Large Velocity Gradient) modeling. 

\item The global SFR is $\simeq$ 1\,$M_\odot$~yr$^{-1}$ based on Spitzer mid-IR data (and GALEX FUV data), also consistent with previous estimates based on IRAS data. Therefore the global star formation efficiency (SFE) is 0.6\,Gyr$^{-1}$ (depletion time $\tau =$ 1.5\,Gyr), similar to that in star forming galaxies \citep[$\sim$ 1\,Gyr, e.g.][]{2018ApJ...853..179T}.

\item The molecular gas and SFR surface densities at 40\,pc scale range from $\Sigma_{\rm mol}$ = 4 to 10$^{3}$\,M$_\sun$\, pc$^{-2}$ and $\Sigma_{\rm SFR}$ = 10$^{-2}$ to 10$^{-0.7}$\,M$_\sun$\, yr$^{-1}$\,kpc$^{-2}$, similar to the typical ranges in nearby disk galaxies.  { Although the scatter is large, the slope of the orthogonal distance regression fit is $N$ = 0.63 (intercept is --2.4) and there is a trend of decreasing slope for regions at inner radii. For the CND $N$ = 0.28, and the intercept is --1.6. }

\item In the pixel-to-pixel (40\,pc) analysis we see that the mean/median SFEs (and standard deviations) for the different regions increase as a function of radius. The CND presents on average a lower SFE (0.3\,Gyr$^{-1}$) by a factor of four than that in the outskirts of the molecular disk.  This is compatible with the range (0.4 -- 1.6\,Gyr$^{-1}$) observed in nearby disk galaxies on 120\,pc scales \citep{2018ApJ...861L..18U}.

\item The global SFE is in agreement with those found for other early-type galaxies in the literature that show similar SFEs to disk galaxies \citep[e.g.][]{2017A&A...605A..74K}. However, this is not consistent with studies reporting lower SFEs for a sample of bulge-dominated galaxies with large dust lanes and with signs of a recent minor merger, properties that match those of Cen~A \citep[e.g.][]{2015MNRAS.449.3503D,2018MNRAS.476..122V}. Cen~A shows that not in all cases is the SFE suppressed in early-type galaxies after gas-rich minor mergers. However, regions with low SFE can also be found within the molecular disk of Cen\,A. The best example is the CND of Cen~A where strong shear and shocks  together with AGN activity \citep{2017ApJ...843..136E} are likely important mechanisms to prevent SF.

\item Giant ellipticals ($M_\star$ $\geq$ 10$^{11}$\,$M_\odot$) like Cen~A with a large amount of molecular gas (likely of external origin), extending $\sim$3\,kpc in radius, and with ongoing SF comparable to that of a spiral galaxy are rare. This can be partly explained because molecular gas can be exhausted by SF in $\sim$1\,Gyr and destroyed by its feedback preferentially in those regions with lower gas densities (i.e. the outskirts of the disk) where self-shielding is limited. This is to be combined with the harsh environmental conditions within the elliptical galaxies, where a hot ISM is present thermally radiating at X-rays.
At a later stage, only dense molecular gas stable against collapse due to dynamical effects will remain, usually in the form of compact components around the center or filamentary structures. These would be the resulting molecular components with low star forming efficiency usually observed in other massive ellipticals.

\item The low SFE in the CND of Centaurus~A is also in contrast to that of low luminosity AGNs and nuclear starbursts, where SFEs are $\sim$1.25 times those of their corresponding disk region \citep{2013AJ....146...19L}. This can be partly a consequence of the high $X_{\rm CO}$ towards the CND of Cen~A (Paper II, \citealt{2014A&A...562A..96I}), which is opposite to the trend observed on average close to the nuclei of AGN/nuclear starbursts  by factors from 2 -- 10 \citep{2013ApJ...777....5S}. Possible causes are stronger shear motions, shocks and the more continuous period of AGN feedback in the case of Cen~A than in  low luminosity AGNs and nuclear starbursts.

\end{enumerate}

\acknowledgments
 We gratefully acknowledge participation in the early stages of this work by T. de Graauw, R. Guesten, E. Loenen, M. Schmalzl and R. Meijerink.
This paper makes use of the following ALMA data:
ADS/JAO.ALMA\#2013.1.00803.S.
ALMA is a partnership of ESO (representing its member states), NSF
(USA) and NINS (Japan), together with NRC (Canada), MOST and ASIAA (Taiwan), and KASI (Republic of
Korea), in cooperation with the Republic of Chile.
The Joint ALMA Observatory is operated by ESO, AUI/NRAO and NAOJ.
The National Radio Astronomy Observatory is a facility of the National Science Foundation operated under cooperative agreement by Associated Universities, Inc.
D.E. was supported by the ALMA Japan Research Grant of NAOJ Chile Observatory, NAOJ-ALMA-0093.
D.E. was supported by JSPS KAKENHI Grant Number JP17K14254.
S.V. acknowledges support by the research projects AYA2014-53506-P and 
AYA2017-84897-P from the Spanish Ministerio de Econom\'{i}a y 
Competitividad, from the European Regional Development Funds (FEDER) and 
the Junta de Andaluc\'{i}a (Spain) grants FQM108. This study  has been 
partially financed by the Consejer\'{i}a de Conocimiento, Investigaci\'{o}n y 
Universidad, Junta de Andaluc\'{i}a and European Regional Development Fund 
(ERDF), ref. SOMM17/6105/UGR. 
Part of this work was achieved using the grant of Visiting Scholars 
Program supported by the Research Coordination Committee, National 
Astronomical Observatory of Japan (NAOJ), National Institutes of Natural 
Sciences (NINS). S.M. would like to thank the Ministry of Science and Technology (MOST)
of Taiwan, MOST 107-2119-M-001-020.
This research has made use of NASA's Astrophysics Data System.
This research made use of Astropy, a community-developed core Python (http://www.python.org) package for Astronomy \citep{2013A&A...558A..33A, 2018arXiv180102634T}; ipython \citep{PER-GRA:2007}; matplotlib \citep{Hunter:2007}; APLpy, an open-source plotting package for Python \citep{2012ascl.soft08017R}; NumPy \citep{2011arXiv1102.1523V}; and scikit-learn, Machine Learning in Python \citep{scikit-learn}.
Data analysis was in part carried out on the open use data analysis computer system at the Astronomy Data Center, ADC, of the National Astronomical Observatory of Japan.
This research has made use of the NASA/ IPAC Infrared Science Archive, which is operated by the Jet Propulsion Laboratory, California Institute of Technology, under contract with the National Aeronautics and Space Administration. 

{\it Facilities:} \facility{ALMA, Spitzer, GALEX, IRSA}

\begin{deluxetable*}{lcrcl}
\tabletypesize{\scriptsize}
\tablecaption{Summary of ALMA CO(1--0) Data\label{tbl1}}
\tablewidth{0pt}
\tablehead{
\colhead{Execution Block IDs} & \colhead{Observation date} & \colhead{Time on Source (min)} & \colhead{Configuration} & Calibrators\tablenotemark{a} \\
}
\startdata
{\bf CO(1--0)} \\\hline
uid://A002/X83dbe6/X9ec   & 2014-06-11 & 27.8 & ext. 12m array  & Ceres, J1427-4206, J1321-4342\\
uid://A002/X95e355/X1ac1 & 2014-12-06 & 13.9 & comp. 12m array & Titan, J1427-4206, J1254-4424\\
uid://A002/X83f101/X165    & 2014-06-12 &  57.2  & 7m array   & Mars, J1427-4206, J1321-4342\\
uid://A002/X83f101/X49a    & 2014-06-11 & 57.2 & " & Mars, J1427-4206, J1321-4342\\
uid://A002/X99b784/Xc88   &  2015-01-16 & 19.4 & TP array &  \\
uid://A002/X99c183/X33fa  &  2015-01-17 & 19.4& "\\
uid://A002/X99c183/X37b2 & 2015-01-17 & 19.4& "\\
uid://A002/X99c183/X3a1e & 2015-01-17 & 19.4& "\\
uid://A002/X9d9aa5/X2058 & 2015-04-08 & 19.4& "\\
uid://A002/X9d9aa5/X23d7 & 2015-04-08 & 19.4& "

\enddata
\tablenotetext{a}{Calibrators: amplitude, bandpass and phase calibrators, in this order. }
\end{deluxetable*}

\begin{deluxetable*}{lcccc}
\tabletypesize{\scriptsize}
\tablecaption{Molecular gas masses and SFEs in different regions of the disk of Centaurus~A \label{tbl3}}
\tablewidth{0pt}
\tablehead{
\colhead{Region\tablenotemark{a}} & \colhead{$M_{\rm mol}$[10$^8$\,$M_\odot$]\tablenotemark{b} } 
& \colhead{mean SFE [Gyr$^{-1}$]} & \colhead{Median SFE [Gyr$^{-1}$]}  & \colhead{ stddev SFE [Gyr$^{-1}$]} 
}
\startdata
CND            & 1.65 &  0.32 & 0.26 & 0.24 \\
Arms           & 2.72 &  0.55 & 0.49 & 0.21 \\
Parallelogram  & 5.99 &  1.02 & 0.88 & 0.52 \\
Outer disk     & 5.65 &  1.29 & 1.05 & 0.93 \\
\enddata
\tablenotetext{a}{Regions as indicated in Fig.\,\ref{figsfr}. Note that in the Circumnuclear Disk (CND), the inner r$<$ 86\,pc is not included due to contamination in the maps by the AGN.}
\tablenotetext{b}{Molecular gas mass assuming $X_{\rm CO}$
  $= N_{\rm H_2} /I _{\rm CO}$ = 2 $\times$ 10$^{20}$~cm$^{-2}$~(K~km~s$^{-1}$)$^{-1}$. For the CND we use
  $X_{\rm CO}$ = 5 $\times$ 10$^{20}$~cm$^{-2}$~(K~km~s$^{-1}$)$^{-1}$ \citep[Paper II, see also][]{2014A&A...562A..96I}.}

\end{deluxetable*}

\clearpage

\bibliography{cenaGMCs.bib,cenaSFR.bib,FlyBy.bib,cenA_8um.bib}

\begin{thebibliography}{}
\expandafter\ifx\csname natexlab\endcsname\relax\def\natexlab#1{#1}\fi

\bibitem[{{Alonso-Herrero} {et~al.}(2006){Alonso-Herrero}, {Rieke}, {Rieke},
  {Colina}, {P{\'e}rez-Gonz{\'a}lez}, \& {Ryder}}]{2006ApJ...650..835A}
{Alonso-Herrero}, A., {Rieke}, G.~H., {Rieke}, M.~J., {et~al.} 2006, \apj, 650,
  835

\bibitem[{{Astropy Collaboration} {et~al.}(2013){Astropy Collaboration},
  {Robitaille}, {Tollerud}, {Greenfield}, {Droettboom}, {Bray}, {Aldcroft},
  {Davis}, {Ginsburg}, {Price-Whelan}, {Kerzendorf}, {Conley}, {Crighton},
  {Barbary}, {Muna}, {Ferguson}, {Grollier}, {Parikh}, {Nair}, {Unther},
  {Deil}, {Woillez}, {Conseil}, {Kramer}, {Turner}, {Singer}, {Fox}, {Weaver},
  {Zabalza}, {Edwards}, {Azalee Bostroem}, {Burke}, {Casey}, {Crawford},
  {Dencheva}, {Ely}, {Jenness}, {Labrie}, {Lim}, {Pierfederici}, {Pontzen},
  {Ptak}, {Refsdal}, {Servillat}, \& {Streicher}}]{2013A&A...558A..33A}
{Astropy Collaboration}, {Robitaille}, T.~P., {Tollerud}, E.~J., {et~al.} 2013,
  \aap, 558, doi:10.1051/0004-6361/201322068

\bibitem[{{Bendo} {et~al.}(2012){Bendo}, {Galliano}, \&
  {Madden}}]{2012MNRAS.423..197B}
{Bendo}, G.~J., {Galliano}, F., \& {Madden}, S.~C. 2012, \mnras, 423, 197

\bibitem[{{Bigiel} {et~al.}(2008){Bigiel}, {Leroy}, {Walter}, {Brinks}, {de
  Blok}, {Madore}, \& {Thornley}}]{2008AJ....136.2846B}
{Bigiel}, F., {Leroy}, A., {Walter}, F., {et~al.} 2008, \aj, 136, 2846

\bibitem[{{Boizelle} {et~al.}(2017){Boizelle}, {Barth}, {Darling}, {Baker},
  {Buote}, {Ho}, \& {Walsh}}]{2017ApJ...845..170B}
{Boizelle}, B.~D., {Barth}, A.~J., {Darling}, J., {et~al.} 2017, \apj, 845, 170

\bibitem[{{Bolatto} {et~al.}(2013){Bolatto}, {Wolfire}, \&
  {Leroy}}]{2013ARA&A..51..207B}
{Bolatto}, A.~D., {Wolfire}, M., \& {Leroy}, A.~K. 2013, \araa, 51, 207

\bibitem[{{Boquien} {et~al.}(2010){Boquien}, {Bendo}, {Calzetti}, {Dale},
  {Engelbracht}, {Kennicutt}, {Lee}, {van Zee}, \&
  {Moustakas}}]{2010ApJ...713..626B}
{Boquien}, M., {Bendo}, G., {Calzetti}, D., {et~al.} 2010, \apj, 713, 626

\bibitem[{{Boquien} {et~al.}(2011){Boquien}, {Calzetti}, {Combes}, {Henkel},
  {Israel}, {Kramer}, {Rela{\~n}o}, {Verley}, {van der Werf}, {Xilouris}, \&
  {HERM33ES Team}}]{2011AJ....142..111B}
{Boquien}, M., {Calzetti}, D., {Combes}, F., {et~al.} 2011, \aj, 142, 111

\bibitem[{{Calzetti}(2011)}]{2011EAS....46..133C}
{Calzetti}, D. 2011, in EAS Publications Series, Vol.~46, EAS Publications
  Series, ed. C.~{Joblin} \& A.~G.~G.~M. {Tielens}, 133--141

\bibitem[{{Calzetti}(2013)}]{2013seg..book..419C}
{Calzetti}, D. 2013, {Star Formation Rate Indicators}, ed.
  J.~{Falc{\'o}n-Barroso} \& J.~H. {Knapen}, 419

\bibitem[{{Calzetti} {et~al.}(2005){Calzetti}, {Kennicutt}, {Bianchi},
  {Thilker}, {Dale}, {Engelbracht}, {Leitherer}, {Meyer}, {Sosey}, {Mutchler},
  {Regan}, {Thornley}, {Armus}, {Bendo}, {Boissier}, {Boselli}, {Draine},
  {Gordon}, {Helou}, {Hollenbach}, {Kewley}, {Madore}, {Martin}, {Murphy},
  {Rieke}, {Rieke}, {Roussel}, {Sheth}, {Smith}, {Walter}, {White}, {Yi},
  {Scoville}, {Polletta}, \& {Lindler}}]{2005ApJ...633..871C}
{Calzetti}, D., {Kennicutt}, R.~C., J., {Bianchi}, L., {et~al.} 2005, \apj,
  633, 871

\bibitem[{{Calzetti} {et~al.}(2007){Calzetti}, {Kennicutt}, {Engelbracht},
  {Leitherer}, {Draine}, {Kewley}, {Moustakas}, {Sosey}, {Dale}, {Gordon},
  {Helou}, {Hollenbach}, {Armus}, {Bendo}, {Bot}, {Buckalew}, {Jarrett}, {Li},
  {Meyer}, {Murphy}, {Prescott}, {Regan}, {Rieke}, {Roussel}, {Sheth}, {Smith},
  {Thornley}, \& {Walter}}]{2007ApJ...666..870C}
{Calzetti}, D., {Kennicutt}, R.~C., {Engelbracht}, C.~W., {et~al.} 2007, \apj,
  666, 870

\bibitem[{{Calzetti} {et~al.}(2010){Calzetti}, {Wu}, {Hong}, {Kennicutt},
  {Lee}, {Dale}, {Engelbracht}, {van Zee}, {Draine}, {Hao}, {Gordon},
  {Moustakas}, {Murphy}, {Regan}, {Begum}, {Block}, {Dalcanton}, {Funes}, {Gil
  de Paz}, {Johnson}, {Sakai}, {Skillman}, {Walter}, {Weisz}, {Williams}, \&
  {Wu}}]{2010ApJ...714.1256C}
{Calzetti}, D., {Wu}, S.-Y., {Hong}, S., {et~al.} 2010, \apj, 714, 1256

\bibitem[{{Casasola} {et~al.}(2015){Casasola}, {Hunt}, {Combes}, \&
  {Garc{\'\i}a-Burillo}}]{2015A&A...577A.135C}
{Casasola}, V., {Hunt}, L., {Combes}, F., \& {Garc{\'\i}a-Burillo}, S. 2015,
  \aap, 577, A135

\bibitem[{{Cluver} {et~al.}(2017){Cluver}, {Jarrett}, {Dale}, {Smith},
  {August}, \& {Brown}}]{2017ApJ...850...68C}
{Cluver}, M.~E., {Jarrett}, T.~H., {Dale}, D.~A., {et~al.} 2017, \apj, 850, 68

\bibitem[{{Colbert} {et~al.}(2004){Colbert}, {Heckman}, {Ptak}, {Strickland},
  \& {Weaver}}]{2004ApJ...602..231C}
{Colbert}, E. J.~M., {Heckman}, T.~M., {Ptak}, A.~F., {Strickland}, D.~K., \&
  {Weaver}, K.~A. 2004, \apj, 602, 231

\bibitem[{{Cox}(2000)}]{2000asqu.book.....C}
{Cox}, A.~N. 2000, {Allen's astrophysical quantities}

\bibitem[{{Crocker} {et~al.}(2011){Crocker}, {Bureau}, {Young}, \&
  {Combes}}]{2011MNRAS.410.1197C}
{Crocker}, A.~F., {Bureau}, M., {Young}, L.~M., \& {Combes}, F. 2011, \mnras,
  410, 1197

\bibitem[{{Daddi} {et~al.}(2010){Daddi}, {Elbaz}, {Walter}, {Bournaud},
  {Salmi}, {Carilli}, {Dannerbauer}, {Dickinson}, {Monaco}, \&
  {Riechers}}]{2010ApJ...714L.118D}
{Daddi}, E., {Elbaz}, D., {Walter}, F., {et~al.} 2010, \apj, 714, L118

\bibitem[{{Dale} \& {Helou}(2002)}]{2002ApJ...576..159D}
{Dale}, D.~A., \& {Helou}, G. 2002, \apj, 576, 159

\bibitem[{{Dale} {et~al.}(2005){Dale}, {Bendo}, {Engelbracht}, {Gordon},
  {Regan}, {Armus}, {Cannon}, {Calzetti}, {Draine}, {Helou}, {Joseph},
  {Kennicutt}, {Li}, {Murphy}, {Roussel}, {Walter}, {Hanson}, {Hollenbach},
  {Jarrett}, {Kewley}, {Lamanna}, {Leitherer}, {Meyer}, {Rieke}, {Rieke},
  {Sheth}, {Smith}, \& {Thornley}}]{2005ApJ...633..857D}
{Dale}, D.~A., {Bendo}, G.~J., {Engelbracht}, C.~W., {et~al.} 2005, \apj, 633,
  857

\bibitem[{{Dame} {et~al.}(2001){Dame}, {Hartmann}, \&
  {Thaddeus}}]{2001ApJ...547..792D}
{Dame}, T.~M., {Hartmann}, D., \& {Thaddeus}, P. 2001, \apj, 547, 792

\bibitem[{{Davis} {et~al.}(2014){Davis}, {Young}, {Crocker}, {Bureau}, {Blitz},
  {Alatalo}, {Emsellem}, {Naab}, {Bayet}, {Bois}, {Bournaud}, {Cappellari},
  {Davies}, {de Zeeuw}, {Duc}, {Khochfar}, {Krajnovi{\'c}}, {Kuntschner},
  {McDermid}, {Morganti}, {Oosterloo}, {Sarzi}, {Scott}, {Serra}, \&
  {Weijmans}}]{2014MNRAS.444.3427D}
{Davis}, T.~A., {Young}, L.~M., {Crocker}, A.~F., {et~al.} 2014, \mnras, 444,
  3427

\bibitem[{{Davis} {et~al.}(2015){Davis}, {Rowlands}, {Allison}, {Shabala},
  {Ting}, {Lagos}, {Kaviraj}, {Bourne}, {Dunne}, {Eales}, {Ivison}, {Maddox},
  {Smith}, {Smith}, \& {Temi}}]{2015MNRAS.449.3503D}
{Davis}, T.~A., {Rowlands}, K., {Allison}, J.~R., {et~al.} 2015, \mnras, 449,
  3503

\bibitem[{{Eckart} {et~al.}(1990){Eckart}, {Cameron}, {Rothermel}, {Wild},
  {Zinnecker}, {Rydbeck}, {Olberg}, \& {Wiklind}}]{1990ApJ...363..451E}
{Eckart}, A., {Cameron}, M., {Rothermel}, H., {et~al.} 1990, \apj, 363, 451

\bibitem[{{Elson} {et~al.}(2019){Elson}, {Kam}, {Chemin}, {Carignan}, \&
  {Jarrett}}]{2019MNRAS.483..931E}
{Elson}, E.~C., {Kam}, S.~Z., {Chemin}, L., {Carignan}, C., \& {Jarrett}, T.~H.
  2019, \mnras, 483, 931

\bibitem[{{Engelbracht} {et~al.}(2005){Engelbracht}, {Gordon}, {Rieke},
  {Werner}, {Dale}, \& {Latter}}]{2005ApJ...628L..29E}
{Engelbracht}, C.~W., {Gordon}, K.~D., {Rieke}, G.~H., {et~al.} 2005, \apjl,
  628, L29

\bibitem[{{Espada} {et~al.}(2012){Espada}, {Matsushita}, {Peck}, {Henkel},
  {Israel}, \& {Iono}}]{2012ApJ...756L..10E}
{Espada}, D., {Matsushita}, S., {Peck}, A.~B., {et~al.} 2012, \apjl, 756, L10

\bibitem[{{Espada} {et~al.}(2009){Espada}, {Matsushita}, {Peck}, {Henkel},
  {Iono}, {Israel}, {Muller}, {Petitpas}, {Pihlstr{\"o}m}, {Taylor}, \&
  {Dinh-V-Trung}}]{2009ApJ...695..116E}
{Espada}, D., {Matsushita}, S., {Peck}, A., {et~al.} 2009, \apj, 695, 116

\bibitem[{{Espada} {et~al.}(2010){Espada}, {Peck}, {Matsushita}, {Sakamoto},
  {Henkel}, {Iono}, {Israel}, {Muller}, {Petitpas}, {Pihlstr{\"o}m}, {Taylor},
  \& {Trung}}]{2010ApJ...720..666E}
{Espada}, D., {Peck}, A.~B., {Matsushita}, S., {et~al.} 2010, \apj, 720, 666

\bibitem[{{Espada} {et~al.}(2017){Espada}, {Matsushita}, {Miura}, {Israel},
  {Neumayer}, {Martin}, {Henkel}, {Izumi}, {Iono}, {Aalto}, {Ott}, {Peck},
  {Quillen}, \& {Kohno}}]{2017ApJ...843..136E}
{Espada}, D., {Matsushita}, S., {Miura}, R.~E., {et~al.} 2017, \apj, 843, 136

\bibitem[{{Fazio} {et~al.}(2004){Fazio}, {Hora}, {Allen}, {Ashby}, {Barmby},
  {Deutsch}, {Huang}, {Kleiner}, {Marengo}, {Megeath}, {Melnick}, {Pahre},
  {Patten}, {Polizotti}, {Smith}, {Taylor}, {Wang}, {Willner}, {Hoffmann},
  {Pipher}, {Forrest}, {McMurty}, {McCreight}, {McKelvey}, {McMurray}, {Koch},
  {Moseley}, {Arendt}, {Mentzell}, {Marx}, {Losch}, {Mayman}, {Eichhorn},
  {Krebs}, {Jhabvala}, {Gezari}, {Fixsen}, {Flores}, {Shakoorzadeh}, {Jungo},
  {Hakun}, {Workman}, {Karpati}, {Kichak}, {Whitley}, {Mann}, {Tollestrup},
  {Eisenhardt}, {Stern}, {Gorjian}, {Bhattacharya}, {Carey}, {Nelson},
  {Glaccum}, {Lacy}, {Lowrance}, {Laine}, {Reach}, {Stauffer}, {Surace},
  {Wilson}, {Wright}, {Hoffman}, {Domingo}, \& {Cohen}}]{2004ApJS..154...10F}
{Fazio}, G.~G., {Hora}, J.~L., {Allen}, L.~E., {et~al.} 2004, The Astrophysical
  Journal Supplement Series, 154, 10

\bibitem[{{Fitzpatrick}(1999)}]{1999PASP..111...63F}
{Fitzpatrick}, E.~L. 1999, \pasp, 111, 63

\bibitem[{{Gil de Paz} {et~al.}(2007){Gil de Paz}, {Boissier}, {Madore},
  {Seibert}, {Joe}, {Boselli}, {Wyder}, {Thilker}, {Bianchi}, {Rey}, {Rich},
  {Barlow}, {Conrow}, {Forster}, {Friedman}, {Martin}, {Morrissey}, {Neff},
  {Schiminovich}, {Small}, {Donas}, {Heckman}, {Lee}, {Milliard}, {Szalay}, \&
  {Yi}}]{2007ApJS..173..185G}
{Gil de Paz}, A., {Boissier}, S., {Madore}, B.~F., {et~al.} 2007, \apjs, 173,
  185

\bibitem[{{Gordon} {et~al.}(2005){Gordon}, {Rieke}, {Engelbracht}, {Muzerolle},
  {Stansberry}, {Misselt}, {Morrison}, {Cadien}, {Young}, {Dole}, {Kelly},
  {Alonso-Herrero}, {Egami}, {Su}, {Papovich}, {Smith}, {Hines}, {Rieke},
  {Blaylock}, {P{\'e}rez-Gonz{\'a}lez}, {Le Floc'h}, {Hinz}, {Latter},
  {Hesselroth}, {Frayer}, {Noriega-Crespo}, {Masci}, {Padgett}, {Smylie}, \&
  {Haegel}}]{2005PASP..117..503G}
{Gordon}, K.~D., {Rieke}, G.~H., {Engelbracht}, C.~W., {et~al.} 2005, \pasp,
  117, 503

\bibitem[{{Hall} {et~al.}(2018){Hall}, {Courteau}, {Jarrett}, {Cluver},
  {Meurer}, {Carignan}, \& {Audcent-Ross}}]{2018ApJ...865..154H}
{Hall}, C., {Courteau}, S., {Jarrett}, T., {et~al.} 2018, \apj, 865, 154

\bibitem[{{Hao} {et~al.}(2011){Hao}, {Kennicutt}, {Johnson}, {Calzetti},
  {Dale}, \& {Moustakas}}]{2011ApJ...741..124H}
{Hao}, C.-N., {Kennicutt}, R.~C., {Johnson}, B.~D., {et~al.} 2011, \apj, 741,
  124

\bibitem[{{Harris} {et~al.}(2010){Harris}, {Rejkuba}, \&
  {Harris}}]{2010PASA...27..457H}
{Harris}, G.~L.~H., {Rejkuba}, M., \& {Harris}, W.~E. 2010, \pasa, 27, 457

\bibitem[{{Hawarden} {et~al.}(1993){Hawarden}, {Sandell}, {Matthews},
  {Friberg}, {Watt}, \& {Smith}}]{1993MNRAS.260..844H}
{Hawarden}, T.~G., {Sandell}, G., {Matthews}, H.~E., {et~al.} 1993, \mnras,
  260, 844

\bibitem[{{Helou} {et~al.}(2004){Helou}, {Roussel}, {Appleton}, {Frayer},
  {Stolovy}, {Storrie-Lombardi}, {Hurt}, {Lowrance}, {Makovoz}, {Masci},
  {Surace}, {Gordon}, {Alonso-Herrero}, {Engelbracht}, {Misselt}, {Rieke},
  {Rieke}, {Willner}, {Pahre}, {Ashby}, {Fazio}, \&
  {Smith}}]{2004ApJS..154..253H}
{Helou}, G., {Roussel}, H., {Appleton}, P., {et~al.} 2004, The Astrophysical
  Journal Supplement Series, 154, 253

\bibitem[{{Henkel} \& {Wiklind}(1997)}]{1997SSRv...81....1H}
{Henkel}, C., \& {Wiklind}, T. 1997, \ssr, 81, 1

\bibitem[{Hunter(2007)}]{Hunter:2007}
Hunter, J.~D. 2007, Computing In Science \& Engineering, 9, 90

\bibitem[{{Israel} {et~al.}(2017){Israel}, {G{\"u}sten}, {Meijerink},
  {Requena-Torres}, \& {Stutzki}}]{2017A&A...599A..53I}
{Israel}, F.~P., {G{\"u}sten}, R., {Meijerink}, R., {Requena-Torres}, M.~A., \&
  {Stutzki}, J. 2017, \aap, 599, A53

\bibitem[{{Israel} {et~al.}(2014){Israel}, {G{\"u}sten}, {Meijerink}, {Loenen},
  {Requena-Torres}, {Stutzki}, {van der Werf}, {Harris}, {Kramer},
  {Martin-Pintado}, \& {Weiss}}]{2014A&A...562A..96I}
{Israel}, F.~P., {G{\"u}sten}, R., {Meijerink}, R., {et~al.} 2014, \aap, 562,
  A96

\bibitem[{{Kannappan} {et~al.}(2009){Kannappan}, {Guie}, \&
  {Baker}}]{2009AJ....138..579K}
{Kannappan}, S.~J., {Guie}, J.~M., \& {Baker}, A.~J. 2009, \aj, 138, 579

\bibitem[{{Kennicutt}(1998)}]{1998ARA&A..36..189K}
{Kennicutt}, Robert~C., J. 1998, Annual Review of Astronomy and Astrophysics,
  36, 189

\bibitem[{{Kennicutt} {et~al.}(2009){Kennicutt}, {Hao}, {Calzetti},
  {Moustakas}, {Dale}, {Bendo}, {Engelbracht}, {Johnson}, \&
  {Lee}}]{2009ApJ...703.1672K}
{Kennicutt}, Robert~C., J., {Hao}, C.-N., {Calzetti}, D., {et~al.} 2009, \apj,
  703, 1672

\bibitem[{{Kennicutt} \& {Evans}(2012)}]{2012ARA&A..50..531K}
{Kennicutt}, R.~C., \& {Evans}, N.~J. 2012, \araa, 50, 531

\bibitem[{{Kennicutt} {et~al.}(2007){Kennicutt}, {Calzetti}, {Walter}, {Helou},
  {Hollenbach}, {Armus}, {Bendo}, {Dale}, {Draine}, {Engelbracht}, {Gordon},
  {Prescott}, {Regan}, {Thornley}, {Bot}, {Brinks}, {de Blok}, {de Mello},
  {Meyer}, {Moustakas}, {Murphy}, {Sheth}, \& {Smith}}]{2007ApJ...671..333K}
{Kennicutt}, Jr., R.~C., {Calzetti}, D., {Walter}, F., {et~al.} 2007, \apj,
  671, 333

\bibitem[{{Kohno} {et~al.}(2002){Kohno}, {Tosaki}, {Matsushita},
  {Vila-Vila{\'o}}, {Shibatsuka}, \& {Kawabe}}]{2002PASJ...54..541K}
{Kohno}, K., {Tosaki}, T., {Matsushita}, S., {et~al.} 2002, Publications of the
  Astronomical Society of Japan, 54, 541

\bibitem[{{Kokusho} {et~al.}(2017){Kokusho}, {Kaneda}, {Bureau}, {Suzuki},
  {Murata}, {Kondo}, \& {Yamagishi}}]{2017A&A...605A..74K}
{Kokusho}, T., {Kaneda}, H., {Bureau}, M., {et~al.} 2017, \aap, 605, A74

\bibitem[{{Kormendy} \& {Ho}(2013)}]{2013ARA&A..51..511K}
{Kormendy}, J., \& {Ho}, L.~C. 2013, Annual Review of Astronomy and
  Astrophysics, 51, 511

\bibitem[{{Kroupa}(2001)}]{2001MNRAS.322..231K}
{Kroupa}, P. 2001, \mnras, 322, 231

\bibitem[{{Leeuw} {et~al.}(2002){Leeuw}, {Hawarden}, {Matthews}, {Robson}, \&
  {Eckart}}]{2002ApJ...565..131L}
{Leeuw}, L.~L., {Hawarden}, T.~G., {Matthews}, H.~E., {Robson}, E.~I., \&
  {Eckart}, A. 2002, \apj, 565, 131

\bibitem[{{Leroy} {et~al.}(2008){Leroy}, {Walter}, {Brinks}, {Bigiel}, {de
  Blok}, {Madore}, \& {Thornley}}]{2008AJ....136.2782L}
{Leroy}, A.~K., {Walter}, F., {Brinks}, E., {et~al.} 2008, \aj, 136, 2782

\bibitem[{{Leroy} {et~al.}(2013){Leroy}, {Walter}, {Sandstrom}, {Schruba},
  {Munoz-Mateos}, {Bigiel}, {Bolatto}, {Brinks}, {de Blok}, {Meidt}, {Rix},
  {Rosolowsky}, {Schinnerer}, {Schuster}, \& {Usero}}]{2013AJ....146...19L}
{Leroy}, A.~K., {Walter}, F., {Sandstrom}, K., {et~al.} 2013, \aj, 146, 19

\bibitem[{{Li} \& {Draine}(2001)}]{2001ApJ...554..778L}
{Li}, A., \& {Draine}, B.~T. 2001, \apj, 554, 778

\bibitem[{{Lisenfeld} {et~al.}(2011){Lisenfeld}, {Espada}, {Verdes-Montenegro},
  {Kuno}, {Leon}, {Sabater}, {Sato}, {Sulentic}, {Verley}, \&
  {Yun}}]{2011A&A...534A.102L}
{Lisenfeld}, U., {Espada}, D., {Verdes-Montenegro}, L., {et~al.} 2011, \aap,
  534, A102

\bibitem[{{Liszt}(2001)}]{2001A&A...371..865L}
{Liszt}, H. 2001, \aap, 371, 865

\bibitem[{{Liu} {et~al.}(2011){Liu}, {Koda}, {Calzetti}, {Fukuhara}, \&
  {Momose}}]{2011ApJ...735...63L}
{Liu}, G., {Koda}, J., {Calzetti}, D., {Fukuhara}, M., \& {Momose}, R. 2011,
  \apj, 735, 63

\bibitem[{{Longmore} {et~al.}(2013){Longmore}, {Bally}, {Testi}, {Purcell},
  {Walsh}, {Bressert}, {Pestalozzi}, {Molinari}, {Ott}, {Cortese}, {Battersby},
  {Murray}, {Lee}, {Kruijssen}, {Schisano}, \& {Elia}}]{2013MNRAS.429..987L}
{Longmore}, S.~N., {Bally}, J., {Testi}, L., {et~al.} 2013, \mnras, 429, 987

\bibitem[{{Mahajan} {et~al.}(2019){Mahajan}, {Ashby}, {Willner}, {Barmby},
  {Fazio}, {Maragkoudakis}, {Raychaudhury}, \& {Zezas}}]{2019MNRAS.482..560M}
{Mahajan}, S., {Ashby}, M.~L.~N., {Willner}, S.~P., {et~al.} 2019, \mnras, 482,
  560

\bibitem[{{Maragkoudakis} {et~al.}(2017){Maragkoudakis}, {Zezas}, {Ashby}, \&
  {Willner}}]{2017MNRAS.466.1192M}
{Maragkoudakis}, A., {Zezas}, A., {Ashby}, M.~L.~N., \& {Willner}, S.~P. 2017,
  \mnras, 466, 1192

\bibitem[{{Marston} \& {Dickens}(1988)}]{1988A&A...193...27M}
{Marston}, A.~P., \& {Dickens}, R.~J. 1988, \aap, 193, 27

\bibitem[{{Martig} {et~al.}(2013){Martig}, {Crocker}, {Bournaud}, {Emsellem},
  {Gabor}, {Alatalo}, {Blitz}, {Bois}, {Bureau}, {Cappellari}, {Davies},
  {Davis}, {Dekel}, {de Zeeuw}, {Duc}, {Falc{\'o}n-Barroso}, {Khochfar},
  {Krajnovi{\'c}}, {Kuntschner}, {Morganti}, {McDermid}, {Naab}, {Oosterloo},
  {Sarzi}, {Scott}, {Serra}, {Griffin}, {Teyssier}, {Weijmans}, \&
  {Young}}]{2013MNRAS.432.1914M}
{Martig}, M., {Crocker}, A.~F., {Bournaud}, F., {et~al.} 2013, \mnras, 432,
  1914

\bibitem[{{Martin} {et~al.}(2005){Martin}, {Fanson}, {Schiminovich},
  {Morrissey}, {Friedman}, {Barlow}, {Conrow}, {Grange}, {Jelinsky},
  {Milliard}, {Siegmund}, {Bianchi}, {Byun}, {Donas}, {Forster}, {Heckman},
  {Lee}, {Madore}, {Malina}, {Neff}, {Rich}, {Small}, {Surber}, {Szalay},
  {Welsh}, \& {Wyder}}]{2005ApJ...619L...1M}
{Martin}, D.~C., {Fanson}, J., {Schiminovich}, D., {et~al.} 2005, \apjl, 619,
  L1

\bibitem[{{Martinez-Badenes} {et~al.}(2012){Martinez-Badenes}, {Lisenfeld},
  {Espada}, {Verdes-Montenegro}, {Garc{\'\i}a-Burillo}, {Leon}, {Sulentic}, \&
  {Yun}}]{2012A&A...540A..96M}
{Martinez-Badenes}, V., {Lisenfeld}, U., {Espada}, D., {et~al.} 2012, \aap,
  540, A96

\bibitem[{{McCoy} {et~al.}(2017){McCoy}, {Ott}, {Meier}, {Muller}, {Espada},
  {Mart{\'\i}n}, {Israel}, {Henkel}, {Impellizzeri}, {Aalto}, {Edwards},
  {Brunthaler}, {Neumayer}, {Peck}, {van der Werf}, \&
  {Feain}}]{2017ApJ...851...76M}
{McCoy}, M., {Ott}, J., {Meier}, D.~S., {et~al.} 2017, \apj, 851, 76

\bibitem[{{McMullin} {et~al.}(2007){McMullin}, {Waters}, {Schiebel}, {Young},
  \& {Golap}}]{2007ASPC..376..127M}
{McMullin}, J.~P., {Waters}, B., {Schiebel}, D., {Young}, W., \& {Golap}, K.
  2007, in Astronomical Society of the Pacific Conference Series, Vol. 376,
  Astronomical Data Analysis Software and Systems XVI, ed. R.~A. {Shaw},
  F.~{Hill}, \& D.~J. {Bell}, 127

\bibitem[{{Mirabel} {et~al.}(1999){Mirabel}, {Laurent}, {Sanders}, {Sauvage},
  {Tagger}, {Charmandaris}, {Vigroux}, {Gallais}, {Cesarsky}, \&
  {Block}}]{1999A&A...341..667M}
{Mirabel}, I.~F., {Laurent}, O., {Sanders}, D.~B., {et~al.} 1999, \aap, 341,
  667

\bibitem[{{Miura} {et~al.}(2019){Miura}, {Espada}, \& {et al}}]{PaperII}
{Miura}, R.~E., {Espada}, D., \& {et al}. 2019, \apj, (Paper II)

\bibitem[{{Miura} {et~al.}(2018){Miura}, {Espada}, {Hirota}, {Nakanishi},
  {Bendo}, \& {Sugai}}]{2018ApJ...864..120M}
{Miura}, R.~E., {Espada}, D., {Hirota}, A., {et~al.} 2018, \apj, 864, 120

\bibitem[{{Miura} {et~al.}(2014){Miura}, {Kohno}, {Tosaki}, {Espada}, {Hirota},
  {Komugi}, {Okumura}, {Kuno}, {Muraoka}, {Onodera}, {Nakanishi}, {Sawada},
  {Kaneko}, {Minamidani}, {Fujii}, \& {Kawabe}}]{2014ApJ...788..167M}
{Miura}, R.~E., {Kohno}, K., {Tosaki}, T., {et~al.} 2014, \apj, 788, 167

\bibitem[{{Muraoka} {et~al.}(2019){Muraoka}, {Sorai}, {Miyamoto}, {Yoda},
  {Morokuma-Matsui}, {Kobayashi}, {Kuroda}, {Kaneko}, {Kuno}, {Takeuchi},
  {Nakanishi}, {Watanabe}, {Tanaka}, {Yasuda}, {Yajima}, {Shibata}, {Salak},
  {Espada}, {Matsumoto}, {Noma}, {Kita}, {Komatsuzaki}, {Kajikawa}, {Yashima},
  {Pan}, {Oi}, {Seta}, \& {Nakai}}]{2019PASJ..tmp...27M}
{Muraoka}, K., {Sorai}, K., {Miyamoto}, Y., {et~al.} 2019, Publications of the
  Astronomical Society of Japan, 27

\bibitem[{{Nicholson} {et~al.}(1992){Nicholson}, {Bland-Hawthorn}, \&
  {Taylor}}]{1992ApJ...387..503N}
{Nicholson}, R.~A., {Bland-Hawthorn}, J., \& {Taylor}, K. 1992, \apj, 387, 503

\bibitem[{{Onodera} {et~al.}(2010){Onodera}, {Kuno}, {Tosaki}, {Kohno},
  {Nakanishi}, {Sawada}, {Muraoka}, {Komugi}, {Miura}, {Kaneko}, {Hirota}, \&
  {Kawabe}}]{2010ApJ...722L.127O}
{Onodera}, S., {Kuno}, N., {Tosaki}, T., {et~al.} 2010, \apj, 722, L127

\bibitem[{{Pancoast} {et~al.}(2010){Pancoast}, {Sajina}, {Lacy},
  {Noriega-Crespo}, \& {Rho}}]{2010ApJ...723..530P}
{Pancoast}, A., {Sajina}, A., {Lacy}, M., {Noriega-Crespo}, A., \& {Rho}, J.
  2010, \apj, 723, 530

\bibitem[{{Parkin} {et~al.}(2012){Parkin}, {Wilson}, {Foyle}, {Baes}, {Bendo},
  {Boselli}, {Boquien}, {Cooray}, {Cormier}, {Davies}, {Eales}, {Galametz},
  {Gomez}, {Lebouteiller}, {Madden}, {Mentuch}, {Page}, {Pohlen}, {Remy},
  {Roussel}, {Sauvage}, {Smith}, \& {Spinoglio}}]{2012MNRAS.422.2291P}
{Parkin}, T.~J., {Wilson}, C.~D., {Foyle}, K., {et~al.} 2012, \mnras, 422, 2291

\bibitem[{{Parkin} {et~al.}(2014){Parkin}, {Wilson}, {Schirm}, {Baes},
  {Boquien}, {Boselli}, {Cormier}, {Galametz}, {Karczewski}, {Lebouteiller},
  {De Looze}, {Madden}, {Roussel}, {Smith}, \&
  {Spinoglio}}]{2014ApJ...787...16P}
{Parkin}, T.~J., {Wilson}, C.~D., {Schirm}, M.~R.~P., {et~al.} 2014, \apj, 787,
  16

\bibitem[{Pedregosa {et~al.}(2011)Pedregosa, Varoquaux, Gramfort, Michel,
  Thirion, Grisel, Blondel, Prettenhofer, Weiss, Dubourg, Vanderplas, Passos,
  Cournapeau, Brucher, Perrot, \& Duchesnay}]{scikit-learn}
Pedregosa, F., Varoquaux, G., Gramfort, A., {et~al.} 2011, Journal of Machine
  Learning Research, 12, 2825

\bibitem[{Perez \& Granger(2007)}]{PER-GRA:2007}
Perez, F., \& Granger, B.~E. 2007, Computing in Science and Engineering, 9, 21

\bibitem[{{P{\'e}rez-Gonz{\'a}lez} {et~al.}(2006){P{\'e}rez-Gonz{\'a}lez},
  {Kennicutt}, {Gordon}, {Misselt}, {Gil de Paz}, {Engelbracht}, {Rieke},
  {Bendo}, {Bianchi}, {Boissier}, {Calzetti}, {Dale}, {Draine}, {Jarrett},
  {Hollenbach}, \& {Prescott}}]{2006ApJ...648..987P}
{P{\'e}rez-Gonz{\'a}lez}, P.~G., {Kennicutt}, Robert~C., J., {Gordon}, K.~D.,
  {et~al.} 2006, \apj, 648, 987

\bibitem[{{Phillips} {et~al.}(1987){Phillips}, {Ellison}, {Keene}, {Leighton},
  {Howard}, {Masson}, {Sanders}, {Veidt}, \& {Young}}]{1987ApJ...322L..73P}
{Phillips}, T.~G., {Ellison}, B.~N., {Keene}, J.~B., {et~al.} 1987, \apjl, 322,
  L73

\bibitem[{{Quillen} {et~al.}(2006{\natexlab{a}}){Quillen}, {Bland-Hawthorn},
  {Brookes}, {Werner}, {Smith}, {Stern}, {Keene}, \&
  {Lawrence}}]{2006ApJ...641L..29Q}
{Quillen}, A.~C., {Bland-Hawthorn}, J., {Brookes}, M.~H., {et~al.}
  2006{\natexlab{a}}, \apj, 641, L29

\bibitem[{{Quillen} {et~al.}(2006{\natexlab{b}}){Quillen}, {Brookes}, {Keene},
  {Stern}, {Lawrence}, \& {Werner}}]{2006ApJ...645.1092Q}
{Quillen}, A.~C., {Brookes}, M.~H., {Keene}, J., {et~al.} 2006{\natexlab{b}},
  \apj, 645, 1092

\bibitem[{{Quillen} {et~al.}(1993){Quillen}, {Graham}, \&
  {Frogel}}]{1993ApJ...412..550Q}
{Quillen}, A.~C., {Graham}, J.~R., \& {Frogel}, J.~A. 1993, \apj, 412, 550

\bibitem[{{Quillen} {et~al.}(2010){Quillen}, {Neumayer}, {Oosterloo}, \&
  {Espada}}]{2010PASA...27..396Q}
{Quillen}, A.~C., {Neumayer}, N., {Oosterloo}, T., \& {Espada}, D. 2010, \pasa,
  27, 396

\bibitem[{{Rieke} {et~al.}(2009){Rieke}, {Alonso-Herrero}, {Weiner},
  {P{\'e}rez-Gonz{\'a}lez}, {Blaylock}, {Donley}, \&
  {Marcillac}}]{2009ApJ...692..556R}
{Rieke}, G.~H., {Alonso-Herrero}, A., {Weiner}, B.~J., {et~al.} 2009, \apj,
  692, 556

\bibitem[{{Rieke} {et~al.}(2004){Rieke}, {Young}, {Cadien}, {Engelbracht},
  {Gordon}, {Kelly}, {Low}, {Misselt}, {Morrison}, {Muzerolle}, {Rivlis},
  {Stansberry}, {Beeman}, {Haller}, {Frayer}, {Latter}, {Noriega-Crespo},
  {Padgett}, {Hines}, {Bean}, {Burmester}, {Heim}, {Glenn}, {Ordonez},
  {Schwenker}, {Siewert}, {Strecker}, {Tennant}, {Troeltzsch}, {Unruh},
  {Warden}, {Ade}, {Alonso-Herrero}, {Blaylock}, {Dole}, {Egami}, {Hinz}, {Le
  Floc'h}, {Papovich}, {Perez-Gonzalez}, {Rieke}, {Smith}, {Su}, {Bennett},
  {Henderson}, {Lu}, {Masci}, {Pesenson}, {Rebull}, {Rho}, {Keene}, {Stolovy},
  {Wachter}, {Wheaton}, {Richards}, {Garner}, {Hegge}, {Henderson}, {MacFeely},
  {Michika}, {Miller}, {Neitenbach}, {Winghart}, {Woodruff}, {Arens},
  {Beichman}, {Gaalema}, {Gautier}, {Lada}, {Mould}, {Neugebauer}, \&
  {Stapelfeldt}}]{2004SPIE.5487...50R}
{Rieke}, G.~H., {Young}, E.~T., {Cadien}, J., {et~al.} 2004, in \procspie, Vol.
  5487, Optical, Infrared, and Millimeter Space Telescopes, ed. J.~C. {Mather},
  50--61

\bibitem[{{Robitaille} \& {Bressert}(2012)}]{2012ascl.soft08017R}
{Robitaille}, T., \& {Bressert}, E. 2012, {APLpy: Astronomical Plotting Library
  in Python}, Astrophysics Source Code Library, , , ascl:1208.017

\bibitem[{{Rydbeck} {et~al.}(1993){Rydbeck}, {Wiklind}, {Cameron}, {Wild},
  {Eckart}, {Genzel}, \& {Rothermel}}]{1993A&A...270L..13R}
{Rydbeck}, G., {Wiklind}, T., {Cameron}, M., {et~al.} 1993, \aap, 270, L13

\bibitem[{{Sandstrom} {et~al.}(2013){Sandstrom}, {Leroy}, {Walter}, {Bolatto},
  {Croxall}, {Draine}, {Wilson}, {Wolfire}, {Calzetti}, {Kennicutt}, {Aniano},
  {Donovan Meyer}, {Usero}, {Bigiel}, {Brinks}, {de Blok}, {Crocker}, {Dale},
  {Engelbracht}, {Galametz}, {Groves}, {Hunt}, {Koda}, {Kreckel}, {Linz},
  {Meidt}, {Pellegrini}, {Rix}, {Roussel}, {Schinnerer}, {Schruba}, {Schuster},
  {Skibba}, {van der Laan}, {Appleton}, {Armus}, {Brandl}, {Gordon}, {Hinz},
  {Krause}, {Montiel}, {Sauvage}, {Schmiedeke}, {Smith}, \&
  {Vigroux}}]{2013ApJ...777....5S}
{Sandstrom}, K.~M., {Leroy}, A.~K., {Walter}, F., {et~al.} 2013, \apj, 777, 5

\bibitem[{{Schlafly} \& {Finkbeiner}(2011)}]{2011ApJ...737..103S}
{Schlafly}, E.~F., \& {Finkbeiner}, D.~P. 2011, \apj, 737, 103

\bibitem[{{Schlegel} {et~al.}(1998){Schlegel}, {Finkbeiner}, \&
  {Davis}}]{1998ApJ...500..525S}
{Schlegel}, D.~J., {Finkbeiner}, D.~P., \& {Davis}, M. 1998, \apj, 500, 525

\bibitem[{{Schmidt}(1959)}]{1959ApJ...129..243S}
{Schmidt}, M. 1959, \apj, 129, 243

\bibitem[{{Shapiro} {et~al.}(2010){Shapiro}, {Falc{\'o}n-Barroso}, {van de
  Ven}, {de Zeeuw}, {Sarzi}, {Bacon}, {Bolatto}, {Cappellari}, {Croton},
  {Davies}, {Emsellem}, {Fakhouri}, {Krajnovi{\'c}}, {Kuntschner}, {McDermid},
  {Peletier}, {van den Bosch}, \& {van der Wolk}}]{2010MNRAS.402.2140S}
{Shapiro}, K.~L., {Falc{\'o}n-Barroso}, J., {van de Ven}, G., {et~al.} 2010,
  \mnras, 402, 2140

\bibitem[{{Solomon} \& {Vanden Bout}(2005)}]{2005ARA&A..43..677S}
{Solomon}, P.~M., \& {Vanden Bout}, P.~A. 2005, \araa, 43, 677

\bibitem[{{Struve} {et~al.}(2010){Struve}, {Oosterloo}, {Morganti}, \&
  {Saripalli}}]{2010A&A...515A..67S}
{Struve}, C., {Oosterloo}, T.~A., {Morganti}, R., \& {Saripalli}, L. 2010,
  \aap, 515, A67

\bibitem[{{Tacconi} {et~al.}(2018){Tacconi}, {Genzel}, {Saintonge}, {Combes},
  {Garc{\'\i}a-Burillo}, {Neri}, {Bolatto}, {Contini}, {F{\"o}rster Schreiber},
  {Lilly}, {Lutz}, {Wuyts}, {Accurso}, {Boissier}, {Boone}, {Bouch{\'e}},
  {Bournaud}, {Burkert}, {Carollo}, {Cooper}, {Cox}, {Feruglio}, {Freundlich},
  {Herrera-Camus}, {Juneau}, {Lippa}, {Naab}, {Renzini}, {Salome}, {Sternberg},
  {Tadaki}, {{\"U}bler}, {Walter}, {Weiner}, \& {Weiss}}]{2018ApJ...853..179T}
{Tacconi}, L.~J., {Genzel}, R., {Saintonge}, A., {et~al.} 2018, \apj, 853, 179

\bibitem[{{The Astropy Collaboration} {et~al.}(2018){The Astropy
  Collaboration}, {Price-Whelan}, {Sip{\'{o}}cz}, {G{\"u}nther}, {Lim},
  {Crawford}, {Conseil}, {Shupe}, {Craig}, {Dencheva}, {Ginsburg},
  {VanderPlas}, {Bradley}, {P{\'e}rez- Su{\'a}rez}, {de Val-Borro}, {Aldcroft},
  {Cruz}, {Robitaille}, {Tollerud}, {Ardelean}, {Babej}, {Bachetti}, {Bakanov},
  {Bamford}, {Barentsen}, {Barmby}, {Baumbach}, {Berry}, {Biscani}, {Boquien},
  {Bostroem}, {Bouma}, {Brammer}, {Bray}, {Breytenbach}, {Buddelmeijer},
  {Burke}, {Calderone}, {Cano Rodr{\'\i}guez}, {Cara}, {Cardoso}, {Cheedella},
  {Copin}, {Crichton}, {D{\'A}vella}, {Deil}, {Depagne}, {Dietrich}, {Donath},
  {Droettboom}, {Earl}, {Erben}, {Fabbro}, {Ferreira}, {Finethy}, {Fox},
  {Garrison}, {Gibbons}, {Goldstein}, {Gommers}, {Greco}, {Greenfield},
  {Groener}, {Grollier}, {Hagen}, {Hirst}, {Homeier}, {Horton}, {Hosseinzadeh},
  {Hu}, {Hunkeler}, {Ivezi{\'c}}, {Jain}, {Jenness}, {Kanarek}, {Kendrew},
  {Kern}, {Kerzendorf}, {Khvalko}, {King}, {Kirkby}, {Kulkarni}, {Kumar},
  {Lee}, {Lenz}, {Littlefair}, {Ma}, {Macleod}, {Mastropietro}, {McCully},
  {Montagnac}, {Morris}, {Mueller}, {Mumford}, {Muna}, {Murphy}, {Nelson},
  {Nguyen}, {Ninan}, {N{\"o}the}, {Ogaz}, {Oh}, {Parejko}, {Parley}, {Pascual},
  {Patil}, {Patil}, {Plunkett}, {Prochaska}, {Rastogi}, {Reddy Janga},
  {Sabater}, {Sakurikar}, {Seifert}, {Sherbert}, {Sherwood-Taylor}, {Shih},
  {Sick}, {Silbiger}, {Singanamalla}, {Singer}, {Sladen}, {Sooley},
  {Sornarajah}, {Streicher}, {Teuben}, {Thomas}, {Tremblay}, {Turner},
  {Terr{\'o}n}, {van Kerkwijk}, {de la Vega}, {Watkins}, {Weaver}, {Whitmore},
  {Woillez}, \& {Zabalza}}]{2018arXiv180102634T}
{The Astropy Collaboration}, {Price-Whelan}, A.~M., {Sip{\'{o}}cz}, B.~M.,
  {et~al.} 2018, ArXiv e-prints, arXiv:1801.02634

\bibitem[{{Tomi{\v{c}}i{\'c}} {et~al.}(2019){Tomi{\v{c}}i{\'c}}, {Ho},
  {Kreckel}, {Schinnerer}, {Leroy}, {Groves}, {Sand strom}, {Blanc}, {Jarrett},
  {Thilker}, {Kapala}, \& {McElroy}}]{2019ApJ...873....3T}
{Tomi{\v{c}}i{\'c}}, N., {Ho}, I.~T., {Kreckel}, K., {et~al.} 2019, \apj, 873,
  3

\bibitem[{{Tsai} {et~al.}(2012){Tsai}, {Hwang}, {Matsushita}, {Baker}, \&
  {Espada}}]{2012ApJ...746..129T}
{Tsai}, M., {Hwang}, C.-Y., {Matsushita}, S., {Baker}, A.~J., \& {Espada}, D.
  2012, \apj, 746, 129

\bibitem[{{Utomo} {et~al.}(2018){Utomo}, {Sun}, {Leroy}, {Kruijssen},
  {Schinnerer}, {Schruba}, {Bigiel}, {Blanc}, {Chevance}, {Emsellem},
  {Herrera}, {Hygate}, {Kreckel}, {Ostriker}, {Pety}, {Querejeta},
  {Rosolowsky}, {Sandstrom}, \& {Usero}}]{2018ApJ...861L..18U}
{Utomo}, D., {Sun}, J., {Leroy}, A.~K., {et~al.} 2018, \apj, 861, L18

\bibitem[{{van de Voort} {et~al.}(2018){van de Voort}, {Davis}, {Matsushita},
  {Rowlands}, {Shabala}, {Allison}, {Ting}, {Sansom}, \& {van der
  Werf}}]{2018MNRAS.476..122V}
{van de Voort}, F., {Davis}, T.~A., {Matsushita}, S., {et~al.} 2018, \mnras,
  476, 122

\bibitem[{{Van Der Walt} {et~al.}(2011){Van Der Walt}, {Colbert}, \&
  {Varoquaux}}]{2011arXiv1102.1523V}
{Van Der Walt}, S., {Colbert}, S.~C., \& {Varoquaux}, G. 2011, ArXiv e-prints,
  arXiv:1102.1523

\bibitem[{VanderPlas(2016)}]{vanderplas2016python}
VanderPlas, J. 2016, Python data science handbook: essential tools for working
  with data (" O'Reilly Media, Inc.")

\bibitem[{{Verley} {et~al.}(2009){Verley}, {Corbelli}, {Giovanardi}, \&
  {Hunt}}]{2009A&A...493..453V}
{Verley}, S., {Corbelli}, E., {Giovanardi}, C., \& {Hunt}, L.~K. 2009, \aap,
  493, 453

\bibitem[{{Verley} {et~al.}(2010){Verley}, {Corbelli}, {Giovanardi}, \&
  {Hunt}}]{2010A&A...510A..64V}
---. 2010, \aap, 510, A64

\bibitem[{{Vila-Vilaro} {et~al.}(2019){Vila-Vilaro}, {Espada}, {Cortes},
  {Leon}, {Pompei}, \& {Cepa}}]{2019ApJ...870...39V}
{Vila-Vilaro}, B., {Espada}, D., {Cortes}, P., {et~al.} 2019, \apj, 870, 39

\bibitem[{{Voss} \& {Gilfanov}(2006)}]{2006A&A...447...71V}
{Voss}, R., \& {Gilfanov}, M. 2006, \aap, 447, 71

\bibitem[{{Wei} {et~al.}(2010){Wei}, {Vogel}, {Kannappan}, {Baker}, {Stark}, \&
  {Laine}}]{2010ApJ...725L..62W}
{Wei}, L.~H., {Vogel}, S.~N., {Kannappan}, S.~J., {et~al.} 2010, \apj, 725, L62

\bibitem[{{Werner} {et~al.}(2004){Werner}, {Roellig}, {Low}, {Rieke}, {Rieke},
  {Hoffmann}, {Young}, {Houck}, {Brandl}, {Fazio}, {Hora}, {Gehrz}, {Helou},
  {Soifer}, {Stauffer}, {Keene}, {Eisenhardt}, {Gallagher}, {Gautier}, {Irace},
  {Lawrence}, {Simmons}, {Van Cleve}, {Jura}, {Wright}, \&
  {Cruikshank}}]{2004ApJS..154....1W}
{Werner}, M.~W., {Roellig}, T.~L., {Low}, F.~J., {et~al.} 2004, \apjs, 154, 1

\bibitem[{{Wu} {et~al.}(2005){Wu}, {Cao}, {Hao}, {Liu}, {Wang}, {Xia}, {Deng},
  \& {Young}}]{2005ApJ...632L..79W}
{Wu}, H., {Cao}, C., {Hao}, C.-N., {et~al.} 2005, \apj, 632, L79

\bibitem[{{Yi} {et~al.}(2005){Yi}, {Yoon}, {Kaviraj}, {Deharveng}, {Rich},
  {Salim}, {Boselli}, {Lee}, {Ree}, {Sohn}, {Rey}, {Lee}, {Rhee}, {Bianchi},
  {Byun}, {Donas}, {Friedman}, {Heckman}, {Jelinsky}, {Madore}, {Malina},
  {Martin}, {Milliard}, {Morrissey}, {Neff}, {Schiminovich}, {Siegmund},
  {Small}, {Szalay}, {Jee}, {Kim}, {Barlow}, {Forster}, {Welsh}, \&
  {Wyder}}]{2005ApJ...619L.111Y}
{Yi}, S.~K., {Yoon}, S.~J., {Kaviraj}, S., {et~al.} 2005, \apj, 619, L111

\bibitem[{{Young} {et~al.}(2014{\natexlab{a}}){Young}, {Gronwall}, {Salzer}, \&
  {Rosenberg}}]{2014MNRAS.443.2711Y}
{Young}, J.~E., {Gronwall}, C., {Salzer}, J.~J., \& {Rosenberg}, J.~L.
  2014{\natexlab{a}}, \mnras, 443, 2711

\bibitem[{{Young} {et~al.}(2009){Young}, {Bendo}, \&
  {Lucero}}]{2009AJ....137.3053Y}
{Young}, L.~M., {Bendo}, G.~J., \& {Lucero}, D.~M. 2009, \aj, 137, 3053

\bibitem[{{Young} {et~al.}(2014{\natexlab{b}}){Young}, {Scott}, {Serra},
  {Alatalo}, {Bayet}, {Blitz}, {Bois}, {Bournaud}, {Bureau}, {Crocker},
  {Cappellari}, {Davies}, {Davis}, {de Zeeuw}, {Duc}, {Emsellem}, {Khochfar},
  {Krajnovi{\'c}}, {Kuntschner}, {McDermid}, {Morganti}, {Naab}, {Oosterloo},
  {Sarzi}, \& {Weijmans}}]{2014MNRAS.444.3408Y}
{Young}, L.~M., {Scott}, N., {Serra}, P., {et~al.} 2014{\natexlab{b}}, \mnras,
  444, 3408

\bibitem[{{Zhu} {et~al.}(2008){Zhu}, {Wu}, {Cao}, \&
  {Li}}]{2008ApJ...686..155Z}
{Zhu}, Y.-N., {Wu}, H., {Cao}, C., \& {Li}, H.-N. 2008, \apj, 686, 155

\end{thebibliography}

\end{document}